\documentclass[12pt,preprint]{aastex}
\usepackage{graphicx}
\usepackage{amsmath}
\def\deg{$^{\mathrm o}$~}
\begin{document}
\title{Polarization of Quasars: Resonant Line Scattering in the Broad Absorption Line Region}
\author{ Hui-Yuan Wang, Ting-Gui Wang and Jun-Xian Wang}
\date{}

\email{whywang@mail.ustc.edu.cn}
\affil{Center for Astrophysics, University of Science and Technology of China,
Hefei, 230026, China\\
Joint Institute of Galaxies and Cosmology, USTC and SHAO, CAS
}

\begin{abstract}
Recent works showed that the absorbing material in broad
absorption line (BAL) quasars is optically thick to major resonant
absorption lines. This material may contribute significantly to
the polarization in the absorption lines. In this paper, we
present a detailed study of the resonant line scattering process
using Monte-Carlo method to constrain the optical depth, the
geometry and the kinematics of BAL Region (BALR). By comparing our
results with observed polarized spectra of BAL quasars, we
find: (1) Resonant scattering can produce polarization up to 9\%
at the absorption trough for doublet transitions and up to 20\%
for singlet transitions in radially accelerated flows. To explain
the large polarization degree in the CIV, NV absorption line
troughs detected in a small fraction of BAL QSOs, a nonmonotonic velocity
distribution along the line of sight or/and additional
contribution from the electron scattering region is required. (2)
The rotation of the flow can lead to the rotation of the
polarization position angle (PA) in the line trough. Large
extending angle of BALR is required to produce the observed large PA rotation
in a few BAL QSOs. (3) A large extending
angle of BALR is required to explain a sub-trough in the polarized
flux that was observed in a number of BAL QSOs. (4) The
resonant-scattering can contribute a significant part of NV
emission line in some QSOs, and may give rise to anomalous strong
NV lines in these quasars. (5) The polarized flux and PA rotation
produced by the resonant scattering in non-BAL is uniquely
asymmetric, which may be used to test the presence of BALR in
non-BAL QSOs.
\end{abstract}


\section{Introduction}

Blue-shifted, broad absorption lines (BAL) are observed in the
ultraviolet spectra of about 10-20\% luminous quasars. These lines
are formed in partially ionized outflows with velocities up to 0.1
c. The outflow is likely driven by intensive radiation of the
quasar probably along the equatorial directions to the extension
at least larger than the broad emission line region, and is likely
several 10 parsecs. Disk wind and material evaporating from the
putative dust torus are two plausible scenarios for the origin of
the gas. It is usually believed that BAL region exists in every
quasar, but only covers a small fraction of quasar sky (Weymann et
al. 1991; Reichard et al. 2003; Green et al.  2001; Hamann,
Korista \& Morris 1993, hereafter HKM93). The outflow may carry a
significant fraction of power released by the accretion process
and momentum into the host galaxy of the quasar, so that it will
influence the subsequent evolution of the galaxy. However, in
order to establish its role, we need to understand many properties
of the outflow such as the global covering factor of BAL region,
the column density and velocity field as a factor of the distance
to the continuum sources.

The absorption line profiles are usually quite stable over time
scales of several to ten years, suggesting of smooth outflow
and/or saturation of the UV emission lines. Similar line strength
from ions of very different abundance and strong absorption
detected in soft X-rays supports latter interpretations (Brinkmann
et al. 1999; Wang et al. 1999; Gallagher et al. 2002). The broad
absorption lines have now probably been detected also in X-rays
with much large column densities. If confirmed, this will suggest
that the very high velocity outflow is already there at very close
to the continuum emission region. Efficient acceleration at small
scale is required.

However, census has yet to be reach on a number of critical
issues: (1) The covering factor of BALR is likely a function of
fundamental parameters, such as the black hole mass and the
accretion rate, which may lead to some difference in the
statistical properties of the BAL QSOs and non-BAL QSOs (Boroson
2002), but we still need to find concrete evidence for this and
their potential relation. (2) Whether the outflow is equatorial or
polar is still a matter of controversy. Recent VLBI observations
of a small number of radio loud BAL quasars with equal number of
steep and flat radio spectra reveal only compact structure in most
case (Jiang D.R. et al. in preparation). While based on the radio
variability, Zhou et al. (2006) proposed polar outflows in six
radio loud quasars. It is still unclear that whether radio loud
BAL quasars are special. Hydrodynamic simulation of accretion disk
wind models, however, results in an equatorial outflow. We note
including poloid magnetic field in the accretion disk may change
the simulation results as for the radio jet model (Blandford \&
Payne 1982); (3) Whether the outflow carry significant angular
momentum, i.e., whether the massive disk wind serves as the driver
of the accretion process. (4) There is big concern whether certain
ultraviolet emission lines such as NV, CIV will be significantly
affected by resonant scattering process, such that we need to
revise our metallicity determination in some objects.

Our current knowledge about the BALR is almost exclusively from
either absorption lines or X-ray absorption edge. Unfortunately,
both types of absorption carries only information of the
absorbing gas along the line of sight to the continuum
source, and we have to rely on the statistics of a large sample of BAL
quasars to obtain the average information of the global
properties. Note such information (on the global properties)
is contained in the scattered light,
i.e., the polarized flux.

Broad Absorption Line QSOs (BAL QSOs) are the only highly
polarized population among radio quiet QSOs (e.g., Stockman,
Moore \& Angel 1984). Its optical/UV continuum shows polarization
(e.g., Stockman et al. 1984; Schmidt \& Hines 1999;
Hutsem$\grave{e}$kers \& Lamy 2001), much larger than that of
non-BAL QSOs. The high continuum polarization is believed due to
the electron scattering probably in the BAL region (BALR, Stockman
et al. 1984; Ogle 1997; Wang, Wang, Wang 2005, hereafter Paper I).
In Paper I we also show that if the BALR exists in all QSOs, and
covers around 20\% of the solid angle, the electron scattering in
the BALR can successfully explain the observed continuum
polarization for both BAL QSOs and non-BAL QSOs. 

Observations show that the polarization is even stronger in the
BAL trough than in the continuum. Ogle et al. (1999, hereafter
O99) presented a spectropolarimetric survey of 36 BAL QSOs, and
found:
\begin{itemize}
\item The BAL troughs are usually more polarized than the
continuum, whereas the broad emission lines are less polarized
(also see Cohen et al. 1995; Hines et al. 1995). Deeper BAL
troughs tend to have higher polarization degrees. The polarization
in the trough can be as high as $\sim$20\%.
\item Position angle
(PA) of the polarization in the troughs are quite common, and smaller
rotations across the corresponding emission lines were found in
some objects.
\item In the spectra of the polarized flux, the
absorption line troughs are usually evident but appear shallower
and show various characteristics: the troughs in polarized flux
are more blueshifted than that in the total flux spectrum in some
objects; a sub-trough emerges to the red side of the CIV, SiIV
and/or NV absorption trough in the polarized flux in several
objects, e.g., 0105-0265, 0226-1024, 1413+1143 $\&$ 1333+2840;
occasionally, a boxy absorption trough, similar to that in the
total flux, was also detected (e.g., in 2225-0534 and 1232+1325).
\item Excess polarized flux across the corresponding emission
lines is observed in several objects (see also Lamy \&
Hutsem$\grave{e}$kers 2004).
\end{itemize}

Two processes were proposed to explain the origin of the polarization
in the absorption line troughs. If the BALR does not cover or
only-partially covers the electron scattering region, the leaked
scattered photons will fill the troughs, thus produce high
polarization (Hines \& Wills 1995; Goodrich \& Miller 1995; Ogle
1997; O99; Lamy \& Hutsem$\grave{e}$kers 2004). On the other hand,
Lee \& Blandford (1997, hereafter LB97) showed that resonant
scattering can produce polarization degree as large as 15\% in the
troughs for doublet transitions. Note LB97 does not
calculate the PA rotation for resonant scattering light. Since
then, our knowledge about the column density, and the geometry of BALR
has changed considerably from the X-ray observations as well as
high resolution UV spectroscopy. In particularly, as showed in
Paper I, the electron scattering in the BALR can produce the
continuum polarization. By taking these new information into
consideration, we will explore in detail the polarization properties
of resonantly scattered light, including the polarization degree,
polarized flux and position angle of the polarization for different
models using Monte Carlo method.

This paper is organized as follows. The geometries and dynamics of
the outflow model and the Monte-Carlo method will be described in
\S 2. The results of Monte-Carlo simulation are given separately
for singlet and doublet transitions in \S 3 and \S 4,
respectively. In \S 5, we will compare our results with the
observed polarized spectra of a sample of BAL QSOs (O99) to put
constraints on the geometries and kinematics of the flow.
Resonantly scattered lines are discussed in \S 6.

\section{Models and the Monte-Carlo Method}

In most theoretic models, the UV-absorbing outflow, initially
launched by gas/radiation pressure or magnetically, is accelerated
through radiation pressure (Murray
et al. 1995, hereafter M95; Proga et al. 2000; Everett 2002;
Konigl \& Kartje 1994; de Kool \& Begelman 1995). In this paper,
following M95, we consider axisymmetric outflow with the BALR
shielded by highly ionized material (hereafter shielding gas) in the
inner region. It is proposed that shielding gas prevents the
outflow from being fully ionized, and its existence was supported
by strong soft X-ray absorption at column densities of a few
10$^{23}$ to $>$ 10$^{24}$ cm$^{-2}$ (e.g., Wang et al. 1999,
Green et al. 2001; Gallagher et al. 2002; Clavel et al. 2006).

\begin{figure}
  \epsscale{0.7}\plotone{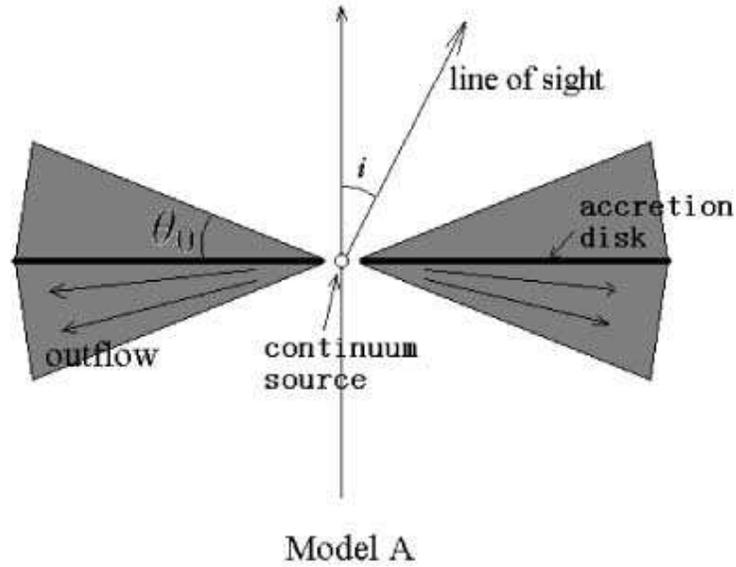}
  \plotone{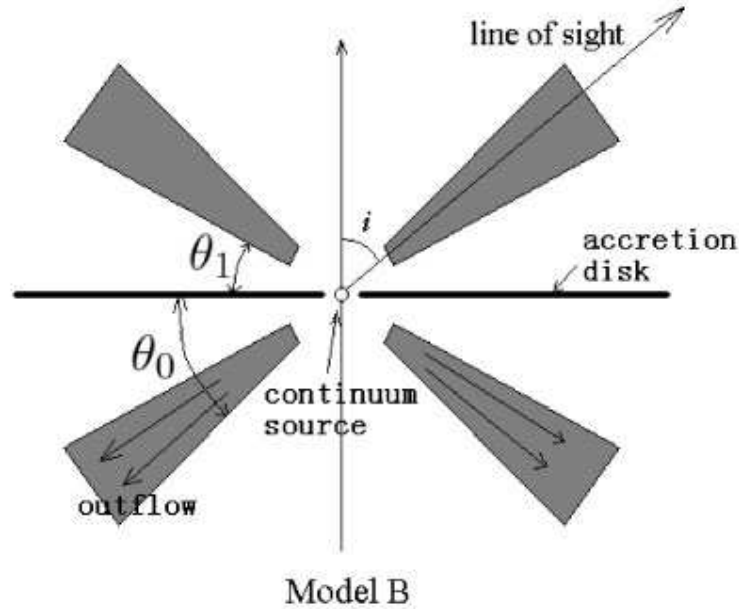}
  \caption{The cross section of the geometry of the outflow used in
   this paper. The thick
   line in the midplane is the accretion disk. The center circle denotes
   isotropic continuum source. The grey region represents the BALR which
   is outflowing. In model A the outflow is equatorial with a half open angle of $\theta_0$ and in model B the outflow
   covers the intermediate inclination angle. $\theta_0$, $\theta_1$ and viewing angle $i$ are defined
   in this figure.}\label{modelab}
\end{figure}

We assume two different geometries for the outflow: model A) the
outflow is equatorial, with a half open angle of $\theta_0$ (see
upper panel of Fig. \ref{modelab}); model B) the axisymmetric
outflow launches at intermediate inclination, covering the
inclination angle between 90$^{\rm o}$$-\theta_0$ to 90$^{\rm
o}$$-\theta_1$ (see lower panel of Fig. \ref{modelab}). Model A
can be considered as a special case of model B with $\theta_1$ =
0. We consider two models of the incident continuum. In the first
model, we assume that the incident continuum is emitted by a point
source, and slightly polarized with $p=1$\%, presumably arising
from a small electron scattering region (hereafter SESR), e.g.,
the shielding gas, around the continuum source (see Paper I).
Unless otherwise specified, the polarization PA is assumed to be
parallel to the symmetry axis of the outflow (PA$_c=$0$^{\rm o}$,
PA$_c$ is the position angle of the polarization in the
continuum), as introduced by scattering of an oblate distribution
of electrons. In the second model, we assume that the electron
scattering region locates just interior to the BALR, and we treat
electron and resonant scattering simultaneously.

By ignoring the velocity in the polar angle direction, we
determine the radial distribution of the density in BALR using the
mass conservation law $n v_r r^2 = const$, where $v_r$ is the
radial velocity and $n$ is the particle density. We assume that the
ionization does not change with the radius (see also LB97 and
M95), thus the density of specific ions also follows $n_i v_r r^2
= const$. This is expected for the dominant ions in the model
of M95, in which the final ionization level is set by the cutoff
of the incident continuum, instead of the ionization parameter.

We adopt M95's description of the radial velocity distribution:
$v_{r}= v_\infty(1-r_{f}/r)^\beta$, where $r_f$ is the launching
radius and
$v_\infty\approx\alpha\sqrt{GM_{BH}/r_f}/(1-\alpha)$(see M95 for
 details) is the terminal radial
velocity. The inner radius of the BALR is set to $r_0$
($r_0>r_f$). Note $\beta= 0.5$ corresponds to a constant ratio of the
radiative force to the gravitational one. The total optical depth
of the resonant scattering at the frequency $\nu$ can be written
as (e.g., Lee 1994)
\begin{equation}
\tau'(x)=\frac{\pi e^2}{m_ec}f_i\frac{c}{\nu_0v_{th}}\phi(x)N_i
\end{equation}
where $f_i$ is the oscillator strength, $\nu_0$ the frequency at
the line center, $v_{th}$ the thermal velocity, $N_i$ the ion
column density, $\phi(x)=1/\sqrt{2\pi}\exp(-x^2/2)$ is the
absorption profile, and
$x=\frac{(\nu-\nu_0)}{\nu_0}\frac{c}{\upsilon_{th}}$.
Notice that the average optical depth over the flow velocity,
\begin{eqnarray}
\tau_0  =  \int\frac{\tau'(x)v_{th}}{v_{\infty}-v_0}dx
        =  \frac{\sqrt{2\pi}v_{th}\tau'(0)}{\Delta v},
\end{eqnarray}
\\
rather than the total optical depth, measures the saturation of
the absorption line, whereas $v_0=v_r(r_0)$ and $\Delta
v=v_{\infty}-v_0$ are the initial velocity and the width
of the absorption trough, respectively.

With these assumptions, the optical depth in the radial direction can
be written as
\begin{eqnarray}
\tau (v_r)  = \frac{\pi e^2}{m_ec}f_i\frac{c}{\nu_0}n_i (\frac{d
v_r}{d r})^{-1}
 = \frac{\Delta
v}{v_r}\frac{\gamma\tau_0}{(v_r/v_0)^{\gamma}-(v_r/v_\infty)^{\gamma}}
\end{eqnarray}
where $\gamma=1-1/\beta$. Note $\beta= 0.5$ produces a constant
optical depth over absorption line profile, or a boxy absorption
line trough. The optical depth decreases for $\beta>0.5$ as velocity
increases, and increases for $\beta<0.5$.

Since the flow is believed to be launched from the accretion disk,
it must also carry angular momentum. Assuming no external torque
and no shearing, we can determine the transverse velocity
$v_{\varphi}$ according to the angular momentum conservation
$rv_\varphi\sin{\theta}=r_fv_{\varphi,0}$, where $v_{\varphi,0}$
is the tangential velocity at the launching radius $r_f$. For a
geometrically thin disk around a super-massive black hole,
$v_{\varphi,0}$$=(GM_{BH}/r_f )^{1/2}$ is the Keplerian velocity
at $r_f$. Note that the $r_f$ and $v_{\varphi,0}$ should be taken
as the radius, and the tangential velocity where the magnetic
torque is no longer important in a magnetically launched wind
(e.g., Konigl \& Kartje 1994; de Kool \& Begelman 1995). We
introduce the rotation parameter $q=v_{\varphi,0}/v_{\infty}$
to describe the rotation of the outflow. In the following simulations,
we will consider four different rotation parameters: 0.0, 0.25, 0.5
and 1.0. Hereafter, we use A$i$-$\theta_0$/B$i$-$\theta_0$-$\theta_1$
to distinguish different models, where $i$ stands for $4q$. For
example, model A1-12$^{\rm o}$ represents model A with a rotation
parameter $q$ =0.25 and $\theta_0=$12$^{\rm o}$, and model B2-45$^{\rm
o}$-30$^{\rm o}$ for model B with $q=0.5$, $\theta_0=$45$^{\rm o}$
and $\theta_1=$30$^{\rm o}$.

Once the outflow model is specified, it is straightforward to
calculate the polarization. Same as Paper I, we denote the density
of the polarized incident photons in direction
$(\theta_i,\varphi_i)$ as
\begin{equation}
\left( \begin{array}{cc}
    \rho_{11}^i &\rho_{12}^i \nonumber \\
    \rho_{21}^i&\rho_{22}^i
    \end{array} \right)
\end{equation}
where $\rho_{ij}$ is the $i$,$j$ component of the photon-density matrix.
The outward density in the direction
$(\theta_o,\varphi_o)$ is
\begin{equation}
\left( \begin{array}{cc}
  \rho_{11}^o & \rho_{12}^o \nonumber \\
  \rho_{21}^o &\rho_{22}^o
\end{array} \right).
\end{equation}

For the transition of singlet, such as the Be-like
CIII$\lambda977$, of which the ground state has $J=0$, we may
write explicitly the outward photon density after one resonant
scattering as $Eq.$ 1-4 of paper I (see also Chandrasekar 1950).
Note most strong BALs are Li-like transitions, e.g., CIV
$\lambda\lambda1548,\;1550$\AA~ and
NV$\lambda\lambda1239,\;1243$\AA. In these cases the ground level
has $J=1/2$ and the two excited levels have either $J=1/2$ (the
lower one) or $J=3/2$ (the upper one). For these transitions the
density of the scattered photon relates to that of the incident
photon by

\begin{equation}
\rho_{\beta\beta'}\propto\sum_{\alpha\alpha'}\sum_{ee'}\sum_{gg'}
\textbf{\emph{M}}_{g'e}(\textbf{\emph{e}}^{out}_{\beta})\textbf{\emph{M}}_{eg}
(\textbf{\emph{e}}^{in}_{\alpha})\rho_{\alpha\alpha'}\textbf{\emph{M}}_{ge'}
(\textbf{\emph{e}}^{in}_{\alpha'})\textbf{\emph{M}}_{e'g'}(\textbf{\emph{e}}^{out}_{\beta'})
\end{equation}
\\
where $\rho_{\beta\beta'}$ and $\rho_{\alpha\alpha'}$ are the
density matrices of the scattered and the incident photons,
respectively; $e$ and $g$ denote the sublevels of the exited and
the ground states; and $\textbf{\emph{M}}$ is defined in $Eq.$ 2.6
of Lee et.al (1994). Following the procedures presented in Paper
I, we calculate the Stokes parameters of the emergent photons for
the given incident radiation and the spatial distribution of ions
with Monte-Carlo simulations. By definition, only photons whose
energy in the rest frame of the ion equals to the difference of
the two energy levels can be resonantly scattered. Doppler shift
due to the bulk motion and thermal motion has to be considered in
the following calculation.

We calculate the four Stokes parameters $I$, $Q$, $U$, $V$ with Eq.
5 of paper I. Other quantities, such as the polarization degree,
polarized flux and PA rotation, which can be observed directly,
are calculated from the Stokes parameters according to the following
formula:

\begin{equation}\label{eq:stokes}
p=\frac{\sqrt{Q^2+U^2}}{I},\;\;f_p=pI,\;\;\tan(2\mathrm{PA})=U/Q
\end{equation}

where $p$, $f_p$ and PA are the linear polarization degree, polarized
flux and polarization position angle, respectively. We do not
consider the circular polarization because no measurement of circular
polarization is available.

\begin{figure}
  \epsscale{1.0}\plotone{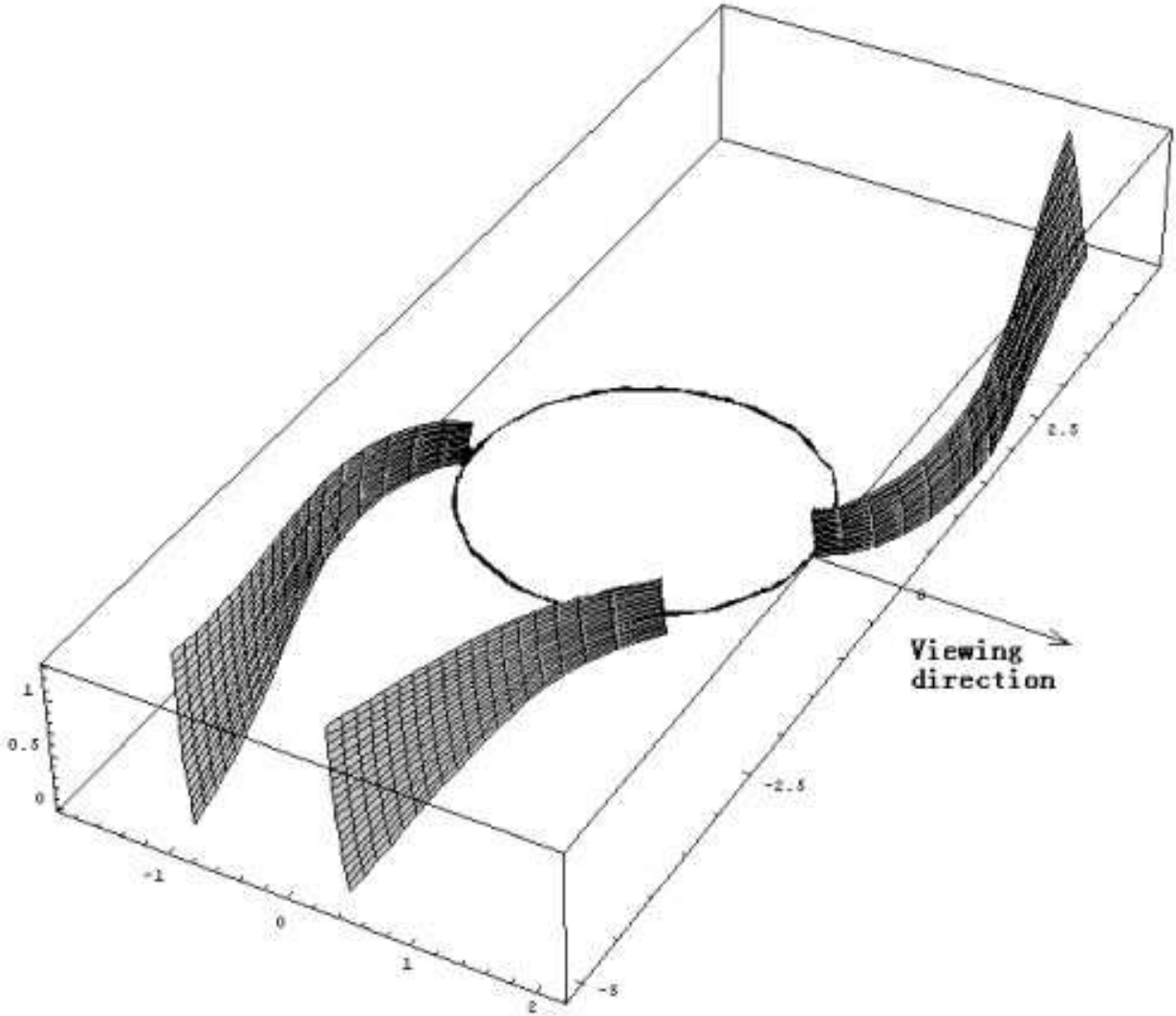}
  \caption{IVS for A-12$^{\rm o}$ model. The photons
   scattered in the three curved surfaces and escaped along the arrow have
   the same observed frequency. Because of the asymmetry of the IVS the PA is rotated.}\label{losurface}
\end{figure}

For the resonant scattering, the optical depth $\tau'$ is
calculated using the Sobolev-Monte Carlo approach (LB97). The
escaped photons are binned in frequency and the polar angle. The
emergent photons with the same frequency in the given direction
come from a curved surface with the same projected flow velocity
along the line of sight (iso-velocity surface, IVS for short), rather
than from the whole volume as in the case of electron scattering. If
the flow has a transverse velocity, such as disk winds, the IVS is
twisted (Fig. \ref{losurface}). This leads to the PA rotation as
we will show below.

Recent works suggested that most common BAL region might be partially covered,
and severely saturated with a typical optical depth $\tau_0>20-80$ for
OVI, CIV, NV.(Hamann 1998; Arav et al. 1999; Wang et al. 2000). Meanwhile
the material might be optically thin for rare elements. In this paper we
will consider a wide range optical depths to include both cases. Since
the polarization of the scattered light for singlet and doublet transitions
is different (e.g., Blandford et al. 2002), we will treat them separately.
We assume $\beta=0.5$ and $\Delta v/v_{\infty} \approx 0.8$ unless
otherwise specified.

\section{Singlet Transitions}

\subsection{Test of the Monte-Carlo Code}

\begin{figure}
\plotone{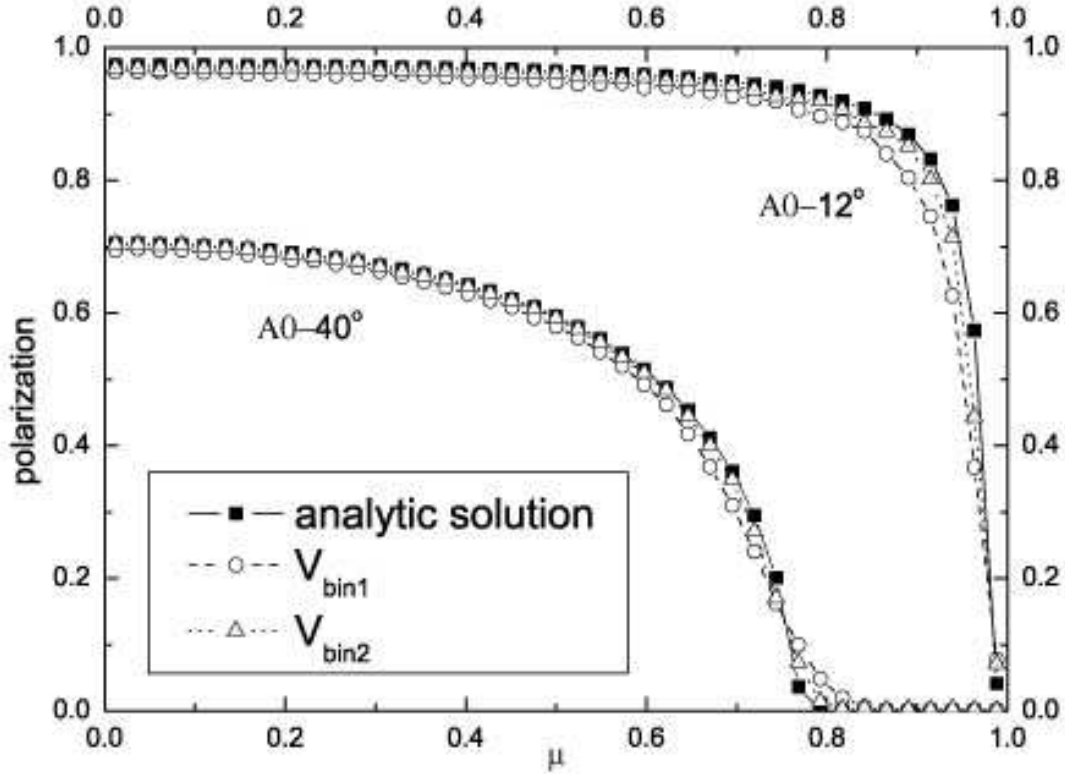} \caption{The polarization of the single scattered
light for singlet transition at line center obtained in the
simulation (open triangles and open circles) and the analytic
solution (black squares) for a pure radial outflow (model
A0-12$^{\rm o}$ and A0-40$^{\rm o}$ and a small
electron-scattering region(SESR) see text for detail). The
triangles and circles are for binsizes of $v_{bin1}=v_\infty/20$
and $v_{bin2}=v_\infty/10$, respectively. The discrepancy between
the analytic solution and the simulation result especially seen
from the polar direction is due to binning
effect.}\label{single-ls}
\end{figure}

In the optically thin limit ($\tau << 1$), the polarization of
resonantly scattered photons can be obtained analytically. By
integrating the scattered photon density over the IVS described in
\S 2, we obtain Stokes parameters at a specific velocity. With the relation
between the scattered and incident photon
density ({\em Eq} 1$\sim$4 of paper I), we obtain the polarization of
the scattered photons at the line center for model A0 (i.e., $q$ =
0) as follows
\[ p=\left\{
           \begin{array}{ll}
               0 &  (i<\theta_0) \\
               \frac{\sin(2\theta_x)}{2\theta_x} & (i>\theta_0),
            \end{array}
            \right.
 \]
\begin{equation}
\theta_x=\arcsin(\frac{\sin\theta_0}{\sin i}),
\end{equation}
\\
where $\theta_0$ and $i$ are defined in Fig. \ref{modelab}. Fig.
\ref{single-ls} shows the results for $\theta_0$=12$^{\rm o}$
(equivalent to a covering factor 20$\%$, as inferred from the
occurrence of BAL QSOs) and $\theta_0=40^{\rm o}$.

This analytic model is used to test our Monte-Carlo code. We run
our Monte-Carlo code for this model, and the result is compared
with the above analytic result (Fig. \ref{single-ls}). The
simulated polarization degree at the line center is consistent
with the analytic one at large inclination angle, but
substantially smaller at low inclinations if we adopt a velocity
bin size $v_\infty/10$. The difference becomes very small for the
bin size $v_\infty/20$. We consider the difference as a
frequency-binning effect, in which the photons in a velocity bin
come from a slice with a finite thickness instead of from an
infinite thin surface, and the average over the IVS slices will
bias toward small polarization. This effect is most significant
when the gradient of polarization with respect to the viewing
angle is largest. Therefore, we adopt a velocity bin size of
$v_\infty/20$ in all the simulations below.

\subsection{Small Electron-Scattering Region (SESR)}

We first consider a polarized incident continuum with $p=1$\%,
presumably arising from an electron scattering region much smaller
than the BALR around the continuum source. The incident continuum
is considered as a point source and only resonant scattered is
taken into account in the simulation. The density and velocity
distribution of the BALR is described in \S2.


\subsubsection{Model A}

\begin{figure}
\plotone{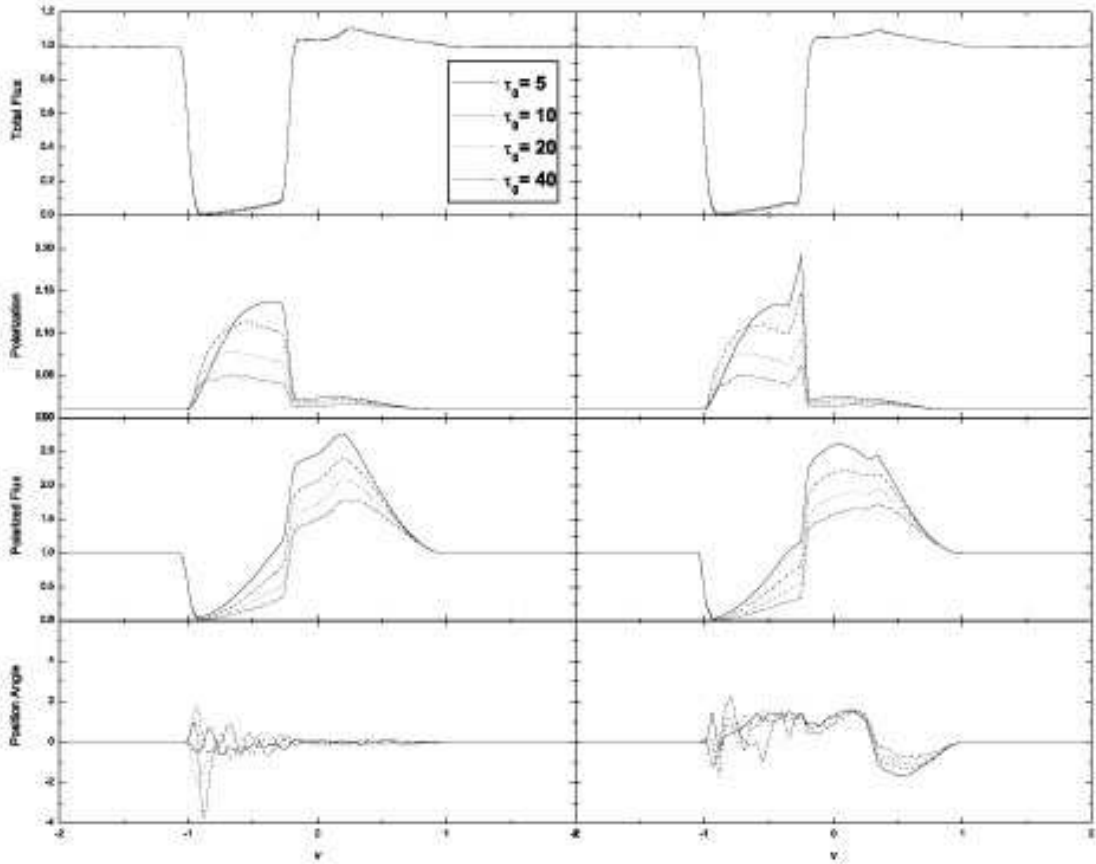} \caption{The average total flux, polarization
degree, polarized flux and
    the PA for singlet in the velocity space for model A-12$^{\rm o}$ and a SESR,
 $\tau_0$=5,10,20,40. $q$ equals to 0 in the left panel and 0.25 in the right panel.
 Here all quantities are averaged over inclination angles for
    BAL QSOs.}\label{a12sne1}
\end{figure}

We summarize the results for a radial outflow (model A0-12$^{\rm
o}$) in the left panel of Fig. \ref{a12sne1}. Obviously, the
absorption line trough has much higher polarization than the
continuum. The polarization degree in the line trough increases
with increasing optical depth for small optical depths $\tau_0$
(see $Eq.$2), reaches a maximum of 14\% at $\tau_0 = 5$, and
declines at larger $\tau_0$. The rising of polarization at small
$\tau_0$ is due to suppression of unpolarized transmitted light
and falling at large $\tau_0$ is due to increasing importance of
multiple scattering, which produces low polarization light. BAL
trough is visible in the polarized flux, but is shallower and
skewed to the blue side than in the total light.
This is caused by the combination of several factors. First of all, the
polarized incident continuum
was resonantly scattered in the BALR, leaving an absorption trough
in the polarized flux. Second, the resonantly scattered line
photons from other directions fill the trough. Note that the
polarization degree of the scattered light is higher at lower
velocities for small and moderate optical depths ($\tau_0=5$)
because the IVS locates closer to the direction
perpendicular to the line of sight\footnote{The angle dependence
of polarization of scattered photons is identical to the Thomson
scattering for singlet (e.g.,Chandrasekhar 1950).}. In addition, there is
much less scattered photons at large velocities because the ions in the
IVS slice becomes smaller as the velocity increases.
To the red side of the absorption trough, there are excess
emission in the polarized flux due to resonantly scattered light,
which has the same PA as the continuum polarization. The peak of
the scattered light is in the red side of emission line around
$v_0$. We note that no rotation of PA relative to the continuum is
produced. The large PA fluctuation at large velocities for large
$\tau_0$'s in Fig. \ref{a12sne1} is due to photon statistics
because only a small number of photons are collected at large
velocity for A0 model.

Next we consider the effect of the rotation of the flow. There is
no direct observational constraint on the value of $q$. However,
if the wind is launched thermally or hydrodynamically from a
Keplerian disk, $v_{\varphi,0}$ is the Keplerian velocity at the
launching radius. With the model describing in \S 2, the starting
site of BALR is close to $r_f$. Since very often the broad
emission lines are also absorbed by BAL, $r_f$ should be no less
than the size of BLR. This sets an upper limit on $v_{\varphi,0}$
to within a factor of two of the broad line width if the latter
is virialized. This suggests $q\sim0.25--0.5$. On the other hand, if
the wind is launched hydromagnetically, then $v_{\varphi,0}$ has
to be redefined as the rotational velocity at the site where
the magnetic torque is no longer important. The velocity should be
larger than the local Keplerian velocity. In the following
analysis, we will consider $q$=0.25, 0.5 and 1.0 (corresponding to
model A1-12$^{\rm o}$, A2-12$^{\rm o}$ and A4-12$^{\rm o}$).

\begin{figure}
\plotone{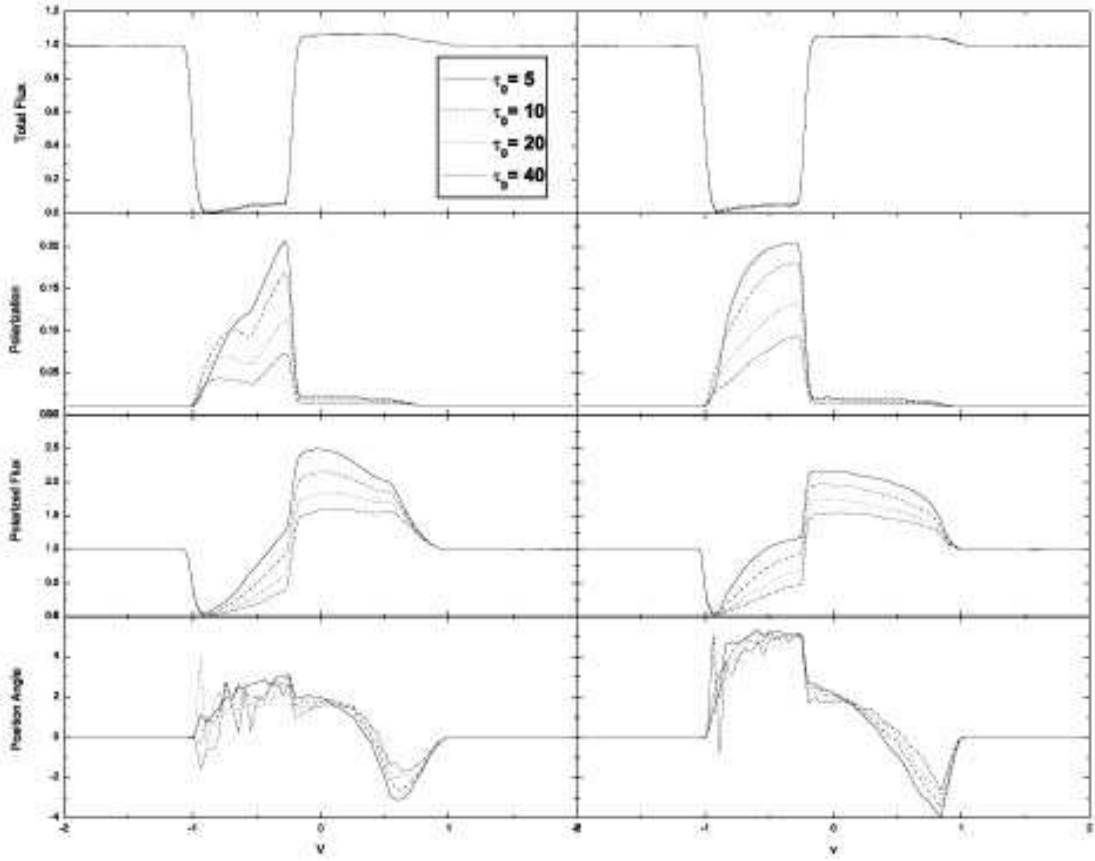}
 \caption{Similar to Fig. \ref{a12sne1} but
for model A2-12$^{\rm o}$(left panel) and A4-12$^{\rm o}$(right
panel), in which flow has a rotational velocity. Rotational of
flow produces a jump in the polarization at around $qv_\infty$ and
PA rotation. }\label{a12sne2}
\end{figure}

The simulated results are summarized in Fig. \ref{a12sne1} and
\ref{a12sne2}. Obviously, the rotational velocity causes a
rotation in PA in the scattered light. The PA rotation is nearly
constant over the absorption trough, but swings from positive to
negative rotation to the red-side of the emission line. PA
rotation in the trough does not depend strongly on the optical
depth. It increases with $q$ because it is sensitive to the
distortion of the IVS, which is determined by the angular
velocity, thus $q$. The rotation also produce a sharp peak in the
polarization between $v_0$ and $qv_\infty$. The peak is caused by
an additional Sobelov surface at projected velocities $<qv_\infty$
(Fig 2.) and an increase in the velocity gradient between
projected velocity $v_0$ and $qv_\infty$. The contribution from
additional Sobelov surface also bring the peak in the polarized
flux very close to $v_0$.

\subsubsection{Model B}

The structure of a quasar might be similar to that of Seyfert
galaxies, in which the line of sight to the equatorial direction
is believed to be blocked by a thick dusty torus (e.g., Antonucci
1993; Dong et al.2005; Mart{\'{\i}}nez-Sansigre et al. 2005). In the
case, BAL QSOs are viewed only at intermediate inclinations, and
the BALR might be confined to intermediate polar angles. Similar
situation arises when the absorbing material is launched vertically
first, and then accelerated radially by the radiation pressure
(see Elvis 2000). In these scenarios, the flow can be approximately
described as model B-$\theta_0$-$\theta_1$ (see Fig. \ref{modelab}).

\begin{figure}
\plotone{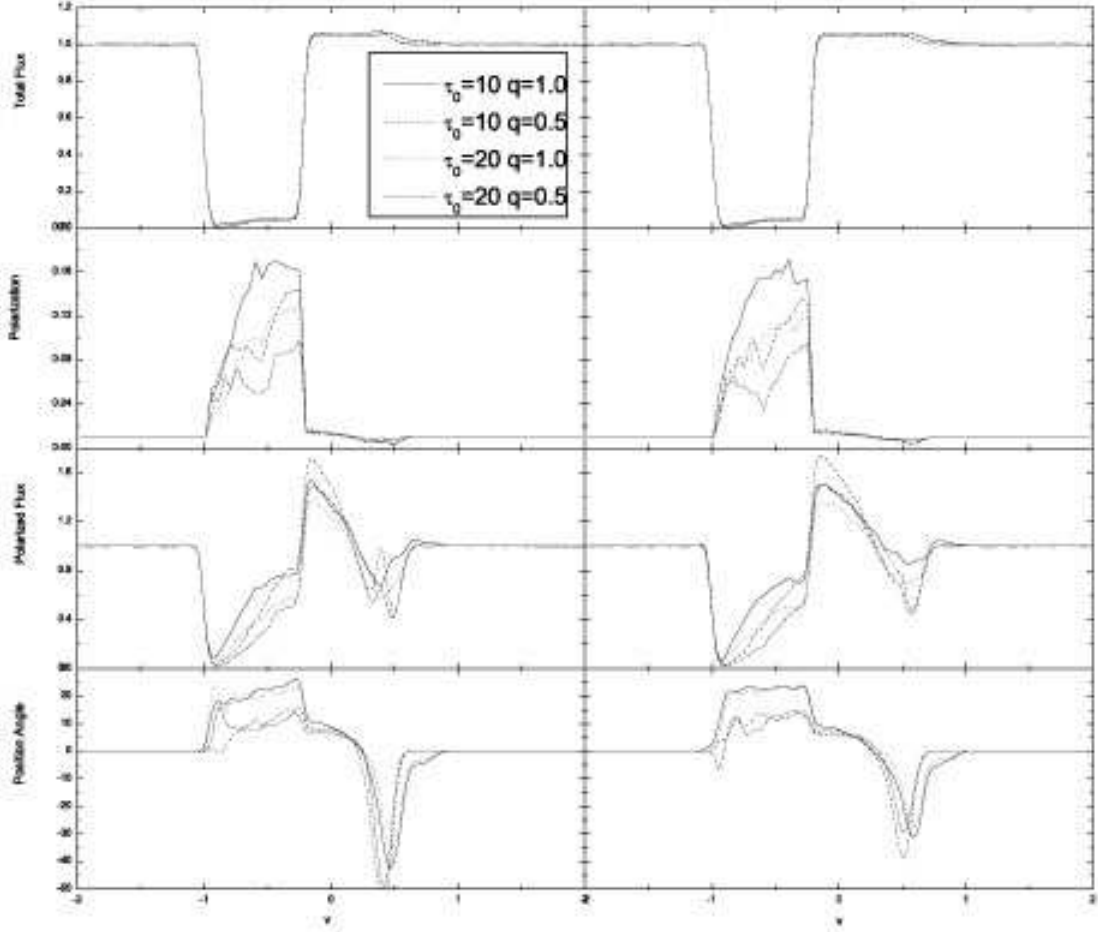}
 \caption{The total flux,
polarization degree, polarized flux and
    the PA for singlet in the velocity space for model B-33$^{\rm o}$-20$^{\rm o}$ and a SESR,
viewing from $i=57^{\rm o}$(left panel) and $i=70^{\rm o}$(right panel
). $\tau_0$ and
     $q$ are marked in the figure.}\label{b33sne}
\end{figure}

\begin{figure}
\plotone{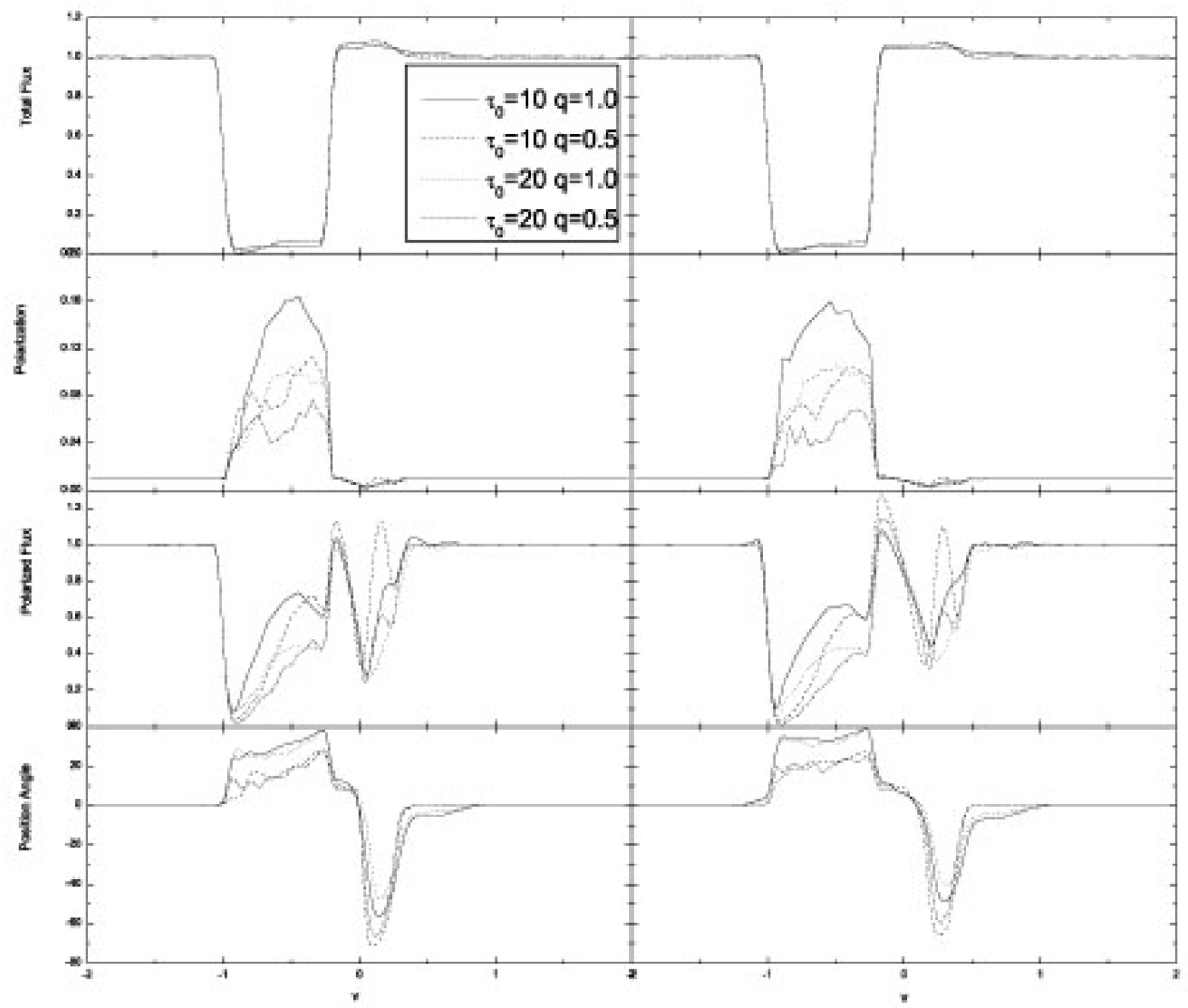}
 \caption{Similar to Fig. \ref{b33sne} but
for model B-45$^{\rm o}$-30$^{\rm o}$ viewing from $i=45^{\rm o}$(left panel) and $i=60^{\rm o}$
(right panel).}\label{b45sne}
\end{figure}

In Figs. \ref{b33sne} and \ref{b45sne}, we show results for model
B-33$^{\rm o}$-20$^{\rm o}$ and B-45$^{\rm o}$-30$^{\rm o}$ with
different $\tau_0$ and $q$. In general, polarization degree and
polarized flux in the absorption line trough looks quite similar
to type-A models: a more highly polarized and blue-skewed trough, a
jump in polarization degree around $qv_\infty$, and PA rotation across
the line profile. There are also several important
differences between the two models. First, type-B models produce
much larger PA rotation in the trough, about 27$^{\rm o}$ for
B4-33$^{\rm o}$-20$^{\rm o}$ and 40$^{\rm o}$ for B4-45$^{\rm
o}$-30$^{\rm o}$. It is easy to understand because
the distortion of the IVS becomes larger due to the angular
momentum conservation. Second, an additional absorption trough
appears to the red side of the emission line in the polarized flux (see
details in Appendix A and discussion below). Subsequently, the
excess of polarized flux around $v_0$ is narrow. The PA rotation
is particular large in the this trough with a negative value.
The resonant scattering light is also more blueshifted than that in
type-A models, as such more scattered photons fill into the trough.

Combining the results of model A and B, we find the polarization
properties (PA, polarized degree, polarized flux and PA) are
sensitive to the rotation parameter $q$ and the geometry of the
model. As in model A, $q$ significantly affects the PA rotation
and the polarization degree in the trough, while the optical depth
alters only the polarization degree and the polarized flux but not
for the PA rotation in the trough.

\subsection{Large Electron-Scattering Region (LESR)}

Now we consider the case that the shielding gas, as the inner part
of the outflow, locates just interior to the inner region of BALR.
In this case, we assume an unpolarized incident continuum, and
simulate the electron scattering in the shielding gas and the
resonant scattering in the BALR outflow simultaneously. Same as in
Paper I, we adopt a constant density model for the shielding gas
with an electron column density of $N_{e} =
4\times10^{23}$~cm$^{-2}$ and the same sky coverage as the BALR
for model A-12$^{\rm o}$, $N_{e} = 3\times10^{23}$~cm$^{-2}$ and
covering factor=$\sin\theta_0$ for model B. With these parameters,
the expected continuum polarization is around 1\%, thus directly
comparable to the SESR models.

\subsubsection{Model A}

\begin{figure}
\plotone{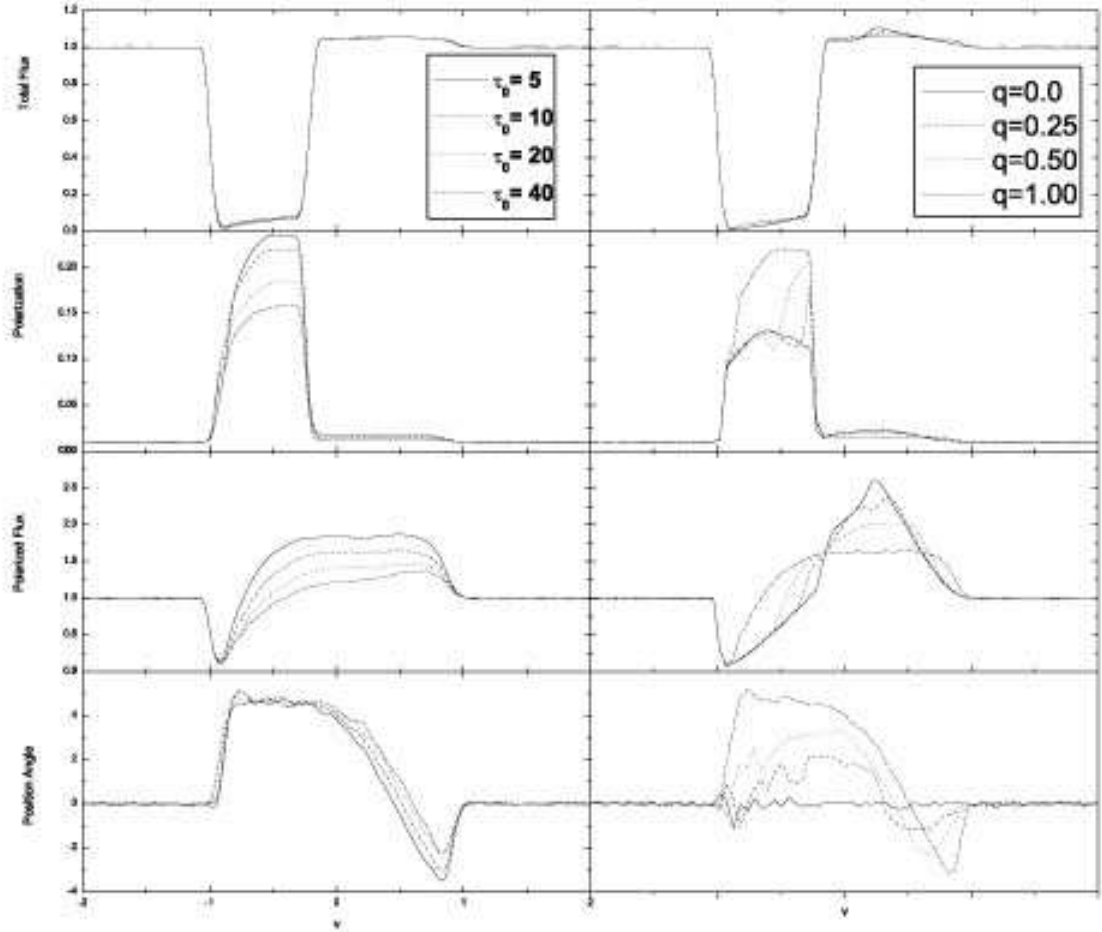} \caption{The total flux, polarization degree,
polarized flux and the PA for singlet in the velocity space for
model A-12$^{\rm o}$ and a large electron-scattering region(LESR)
on the base of the outflow. In the left panel, $q=1.0$ and
different $\tau_0$ is marked. In the right panel $\tau_0=10$ and
different $q$ are marked.}\label{a12se}
\end{figure}

Fig. \ref{a12se} presents the polarization degree, the polarized
flux, PA rotation and the total flux for model A-12$^{\rm o}$. As
expected, the continuum polarization degree is around 1\% and its
position angle is aligned with the symmetric axis. These results
retain some of feature of the corresponding SESR models, such as
excess polarization in the velocity range $-v_{\phi,0}$-$-v_0$, PA
rotations for none-$q$ models. However, we find that the
polarization degree in the trough is higher and decreases more
slowly with increasing $\tau_0$ than in the model A-12$^{\rm o}$
with SESR. We also notice that the absorption trough appears only
to the blue side of $v\simeq -v_{\varphi,0}$ in the polarized
flux, and the PA rotation has no apparent jump around $-v_0$,
which appears in the SESR models. These characteristics are
remarkably different from models with SESR discussed in \S3.1.1.

\begin{figure}
\epsscale{1}\plotone{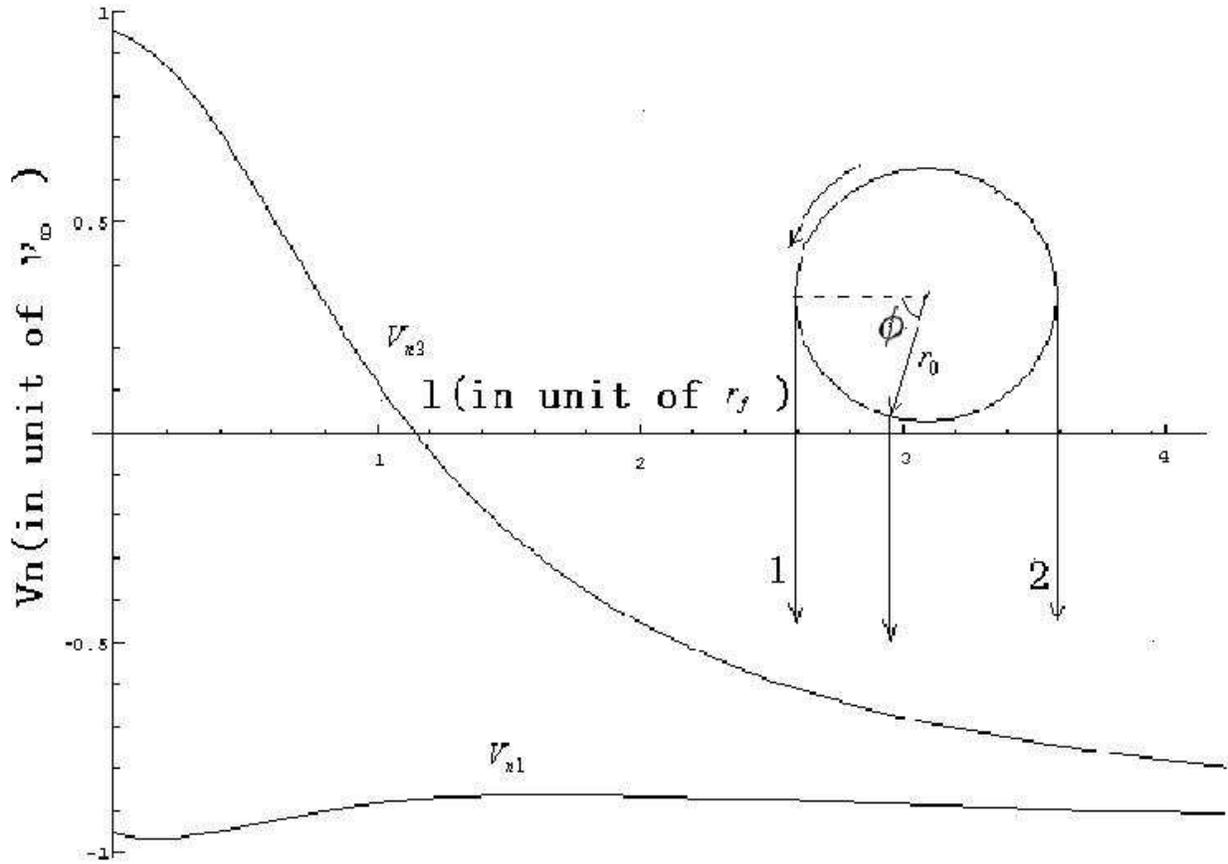} \caption{The projected velocity of
the flow encountered by the photons from two sites towards the
observer. The circle denotes the outer edge of the electron
scattering region or the inner edge of the outflow.
$\upsilon_{n1}$ is the projected velocity on the path 1 and
$\upsilon_{n2}$ on the path 2. Obviously, photons only in a narrow
frequency range are scattered by ions along the path 1, and
photons with a wide range of frequence can be scattered  along the
path 2.  }\label{rotation}
\end{figure}

It should be pointed out that the differences arise because the
'leaked' electron-scattered photons fill the trough. This is
illustrated in Fig. \ref{rotation}, which shows the LOS(line of
sight) velocities of the outflow along the track of the photons
scattered by electrons at two different sites (marked by 1 and 2)
at the equatorial plane. In the figure, the velocity towards us is
taken as negative and away from us as positive. When the rotation
of the flow is considered, photons with $\delta\lambda/ \lambda_0$
in [-$v_{\infty}/c$, $v_{\varphi,0}/c$] may be scattered by ions
along track 2, while only those with $\delta\lambda/\lambda_0$ in
[-$v_{\infty}/c$,-$v_{\varphi,0}/c$ ] may be scattered along track
1, where $\lambda_0$ is the absorption wavelength of the ion in
the rest frame. In other words, photons with
$\delta\lambda/\lambda_0$ in [-$v_{\varphi,0}/c$, $v_{\infty}/c$]
will reach the observer freely along track 1. For a general track
crossing the circle of radius $r_f$ at $\phi$ (defined in Fig.
\ref{rotation}, in the equatorial plane for simplicity), the
scattered continuum in the wavelength range of
$\delta\lambda/\lambda_0$ in [-$v_{\infty}/c$, -$(v_0
\sin\phi$+$v_{\varphi,0}\cos\phi)/c$] will be absorbed. It is
interesting to observe that the starting wavelength of resonant
absorption to the electron scattered photons varies from
$-v_{\varphi,0}$ (at $\varphi=0$) to $-v_0$ (at $\varphi=\pi/2$)
then to $v_{\varphi,0}$ ($\varphi=\pi$) as we look at the electron
scattering sphere from left to right in Fig. \ref{rotation}. Since
the variation is smooth, there is no jump in the polarized light
at $v_0$. In comparison with SESR models, the polarized flux in
the velocity range [-$v_{\varphi,0}$, -$v_0$] is enhanced due to
the contribution of leaked scattered continuum, whereas it is
reduced in the velocity range [-$v_0$, $v_{\varphi,0}$] due to
absorption of scattered light in the left half BALR region. If
rotation velocity is comparable to $v_\infty$, the
electron-scattered light will fill almost the entire trough except
around $-v_\infty$, leaving a narrow absorption trough there in
the polarized light (the left panel of Fig. \ref{a12se}).

\begin{figure}
\plotone{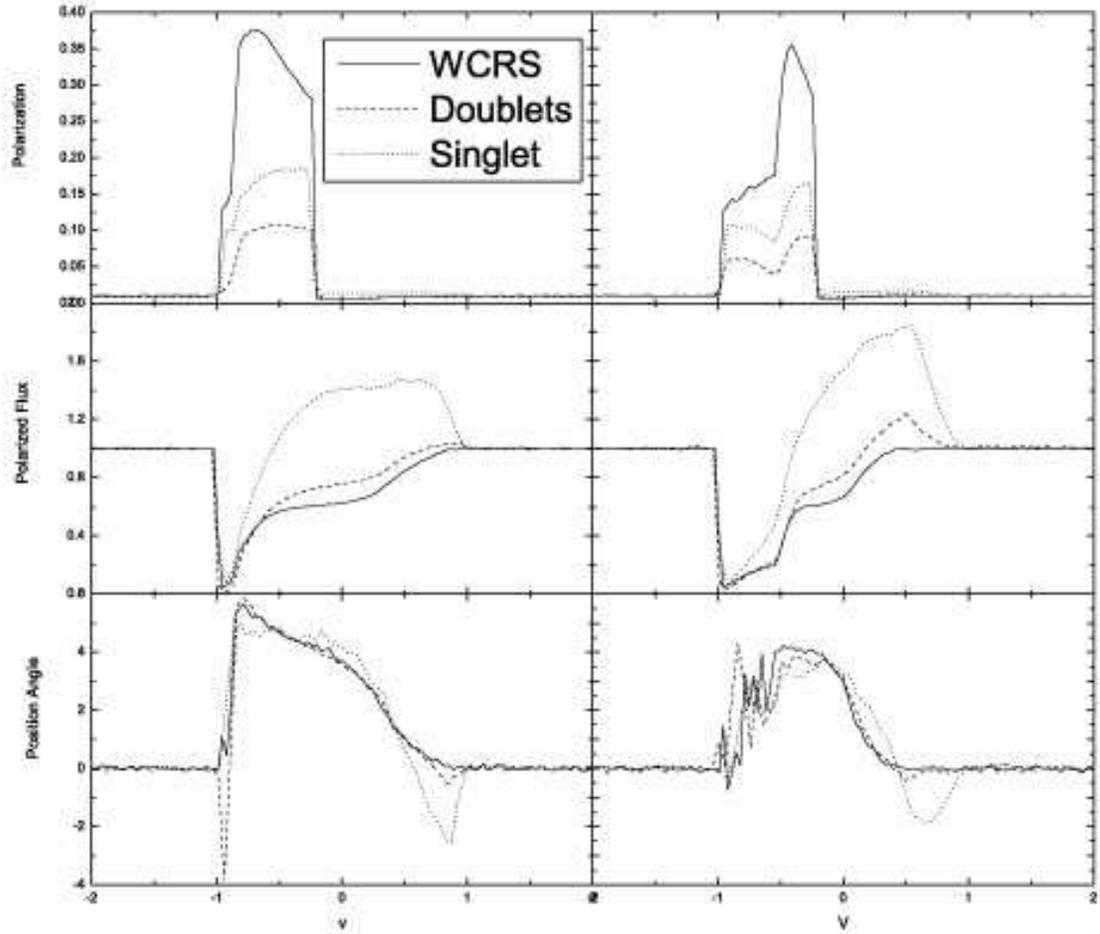} \caption{ The polarization degree, polarized
flux and the PA in velocity space for WCRS, singlet and doublets
model. All are with a LESR. The left panel is for model
A4-12$^{\rm o}$ and right one is for model A2-12$^{\rm o}$.
$\tau_0=20$ for both doublets and singlet. }\label{abcu12}
\end{figure}

To further support this analysis, we plot the polarized flux, its
position angle and polarization degree of electron scattering
light (labelled WCRS, without the contribution of resonant
scattering) in Fig. \ref{abcu12}. It includes the transmitted
light from electron scattering region. The polarized flux rises
almost linearly in the velocity range [$-v_\infty$, $\sim
-v_{\varphi,0}$], presumedly due to variation in the optical
depth, then increases very rapidly around $-v_{\varphi,0}$, reachs
a plateau, and it increases to the continuum level at about
$v_{\infty}$. The position angle of the scattered continuum
rotates across the line profile in a manner similar to the
resonantly scattered light because at a given frequency the leaked
region is no longer rotationally symmetric.

In Fig. \ref{abcu12}, we also show the final polarization degree,
polarized flux, and position angle to illustrate the relative
contribution of the leaked light. It is obvious that the leaked
light is more polarized than resonantly scattered light because of
large line optical depth considered in the model (panel A). The
relative importance of the leaked continuum depends on
the line optical depth, and become more important at large optical
depth because the polarization of resonantly scattered light
decreases with increasing optical depth while that of leaked
continuum does not. As a result the polarization degree is less
sensitive to optical depth in a LESR model than the corresponding
 SESR model. For example, the highest polarization in the trough
drops from 21\% to 8\% (a factor of $\sim 2.6$) when $\tau_0$ increases
from 5 to 40 for A4-12$^{\rm o}$ model with SESR. It changes only 50\%
for the same model with LESR.

At small $q$, such leakage is insignificant. An apparent jump
appears around $v=-v_{\varphi,0}$ in the polarization: the
polarization is larger at $v>-v_{\varphi,0}$ and smaller at
$v<-v_{\varphi,0}$ (Fig. \ref{a12se}). Similar jumps also appear
in the panels of the polarized flux and the PA. These jumps are
caused by the fact that the leakage contributes mostly to the
polarized flux at $v>-v_{\varphi,0}$, whereas resonantly scattered
light works at at both large and small velocities.

\subsubsection{Model B}

\begin{figure}
\plotone{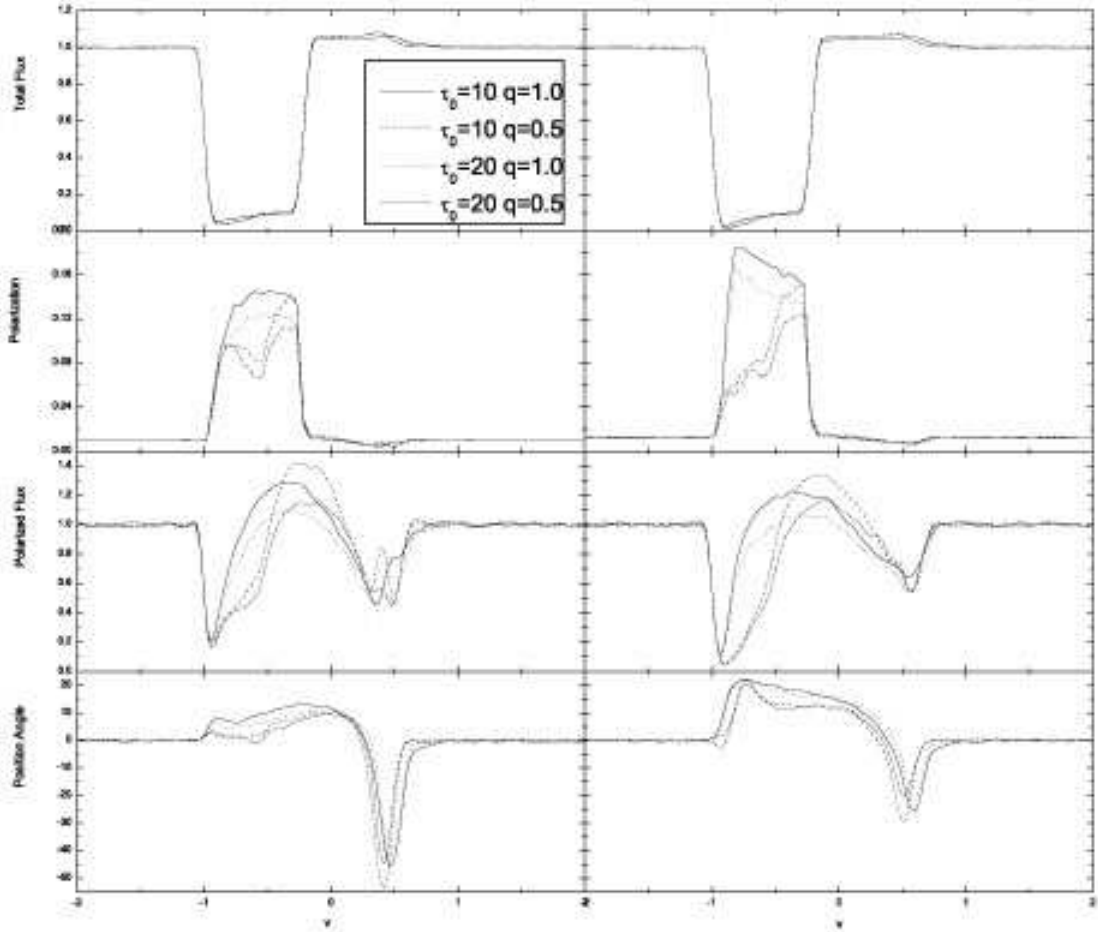} \caption{The total flux, polarization degree,
polarized flux and the PA for singlet in the velocity space for
model B-33$^{\rm o}$-20$^{\rm o}$ and a LESR . The left panel is
viewing from $i=57^{\rm o}$ and the right one is viewing from
$i=70^{\rm o}$. $q$ and $\tau_0$ are marked in the
figure.}\label{b33se}
\end{figure}

\begin{figure}
\plotone{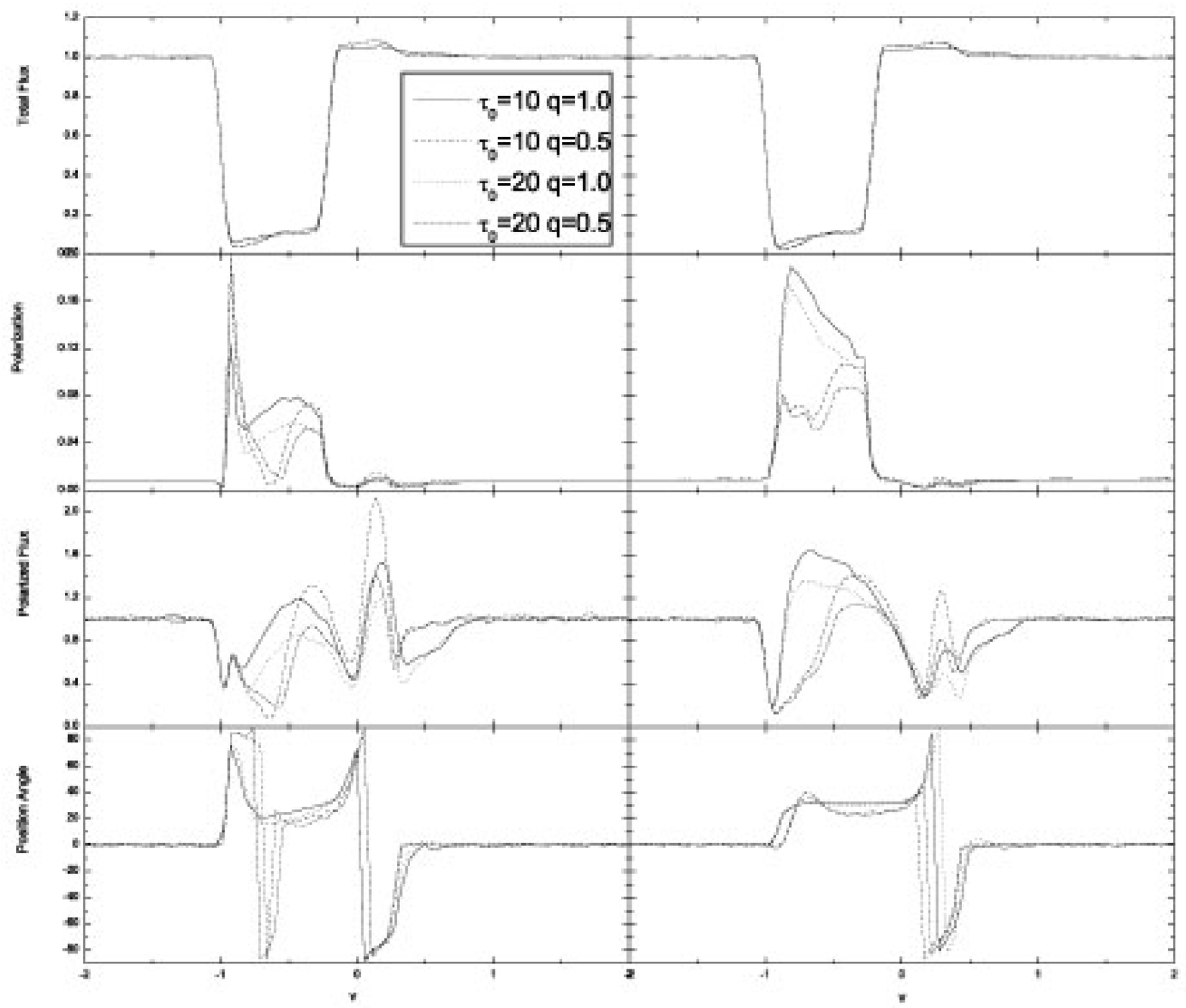} \caption{ Similar to Fig. \ref{b33se} but for
model B-45$^{\rm o}$-30$^{\rm o}$ viewing from $i=45^{\rm o}$(left
panel) and $i=60^{\rm o}$(right panel).}\label{b45se}
\end{figure}

We show the results in Fig. \ref{b33se} (for model B-33$^{\rm o}$-20$^{\rm o}$)
 and \ref{b45se} (for model B-45$^{\rm o}$-30$^{\rm o}$). The effect of LESR
on type-B models is quite similar to that on type-A models, and can be
attributed mainly to the 'leakage' of the
scattered continuum. For instance, the jump in the polarized flux,
PA rotation in the SESR models at $v=-v_0$ is now replaced by a
smooth profile. Polarized flux rises steeply at $v\simeq
-v_{\varphi,0}$ due to the selective leakage.

\begin{figure}
\plotone{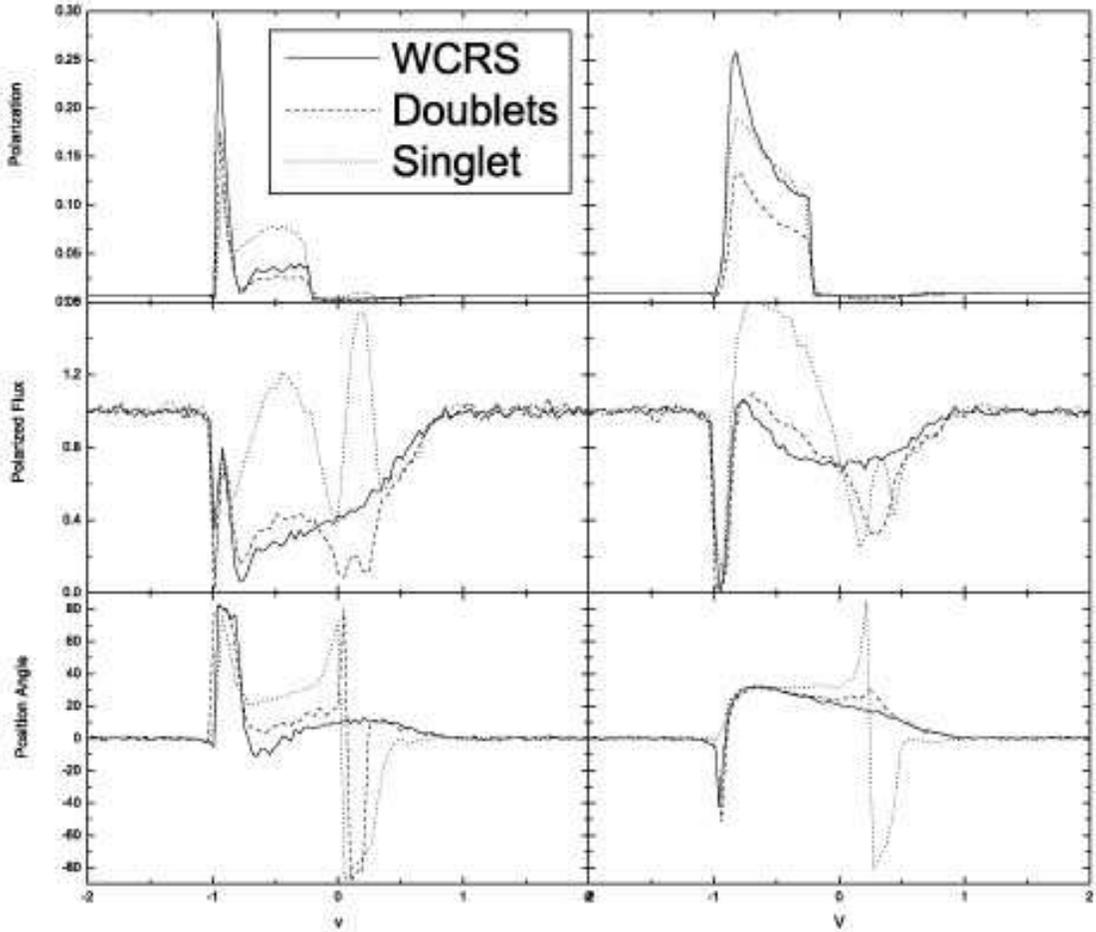} \caption{ The polarization degree, polarized
flux and PA in the velocity space for WCRS, singlet and doublets
model. The model adopted is
 model B4-45$^{\rm o}$-30$^{\rm o}$ and a LESR, viewing from $i=45^{\rm o}$(left panel) and
viewing from $i=60^{\rm o}$(right panel), assuming $\tau_0=10$.
}\label{abcu-b4-45}
\end{figure}

\begin{figure}
\plotone{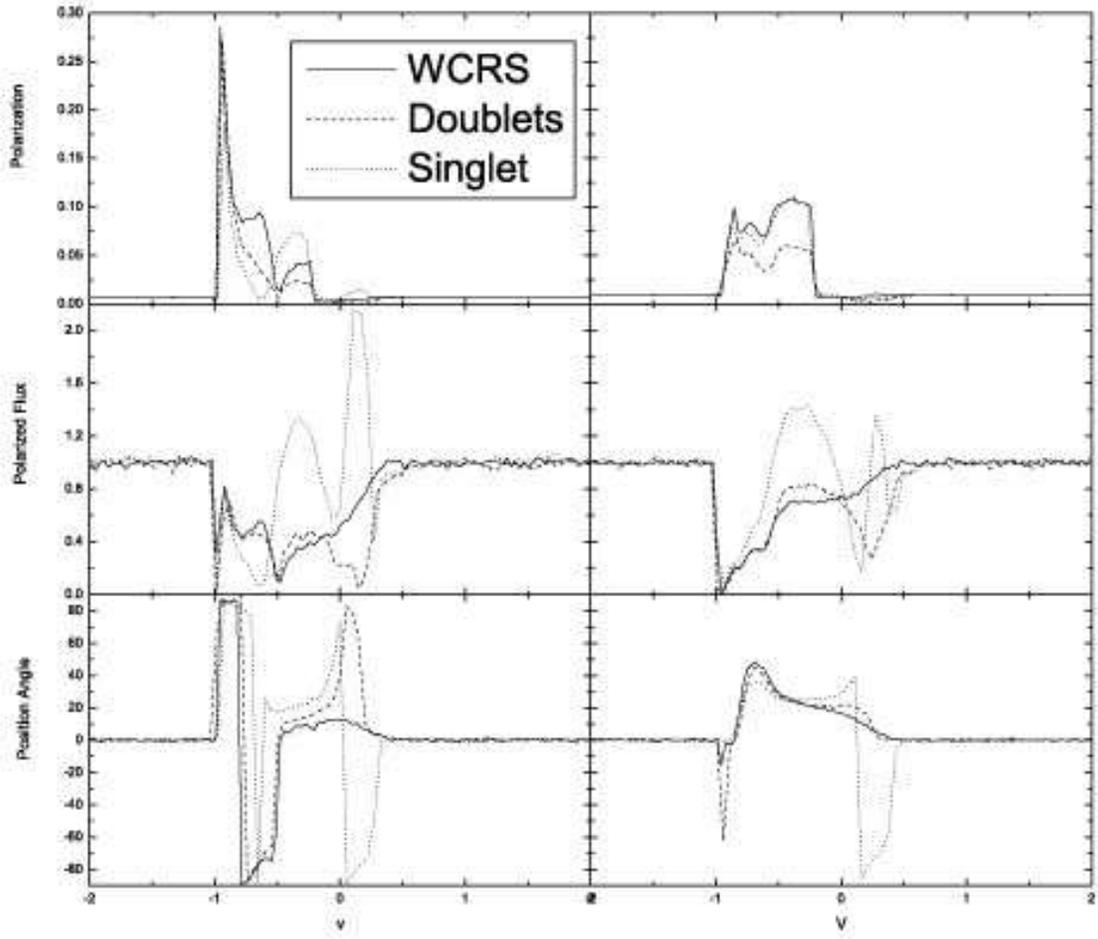} \caption{ Similar to Fig. \ref{abcu-b4-45} but
for model B2-45$^{\rm o}$-30$^{\rm o}$. }\label{abcu-b2-45}
\end{figure}

However, model B-45$^{\rm o}$-30$^{\rm o}$ shows several
distinctive characteristics that do not appear in previous models.
First, PA swings from parallel to nearly perpendicular at $v\simeq
-v_\infty$, and then back to parallel at $v\simeq -0.8v_\infty$ at
viewing angle $i=45$\deg. At large inclinations, the PA rotation
appears smaller. Another PA swing of 90\deg occurs at $v>0$. To
understand the origin of these characteristics we plot the
polarization degree, polarized flux and PA of models for doublet,
singlet and  WCRS in Figs. \ref{abcu-b4-45} and \ref{abcu-b2-45}.
From these figures, we easily find that the blue PA swing is
caused by the electron scattering and the red PA swing is due to
the resonant scattering.

In the previous section we introduce two polarizing processes in
the line trough, resonant scattering and selective leakage. Here,
an additional process, back-scattering of electron in the
shielding gas, is also important for model B, especially at large
$\theta_0$. In model A with small $\theta_0$, most of
back-scattering photons intercept the BALR, whereas in model B
only photons with velocity $\sim -v_{\infty}$  intercept the BALR,
and others do not, thus fill the trough
directly. The polarization, polarized flux and PA rotation are
determined by the competition among the three
processes. In order to investigate the back-scattering we
integrate electron-scattering photon density in the region
with 90$^{\rm o}<\varphi_i-\varphi_o<270^{\rm o}$ according to
$Eq.$ 1-5 of Paper I and obtain the Stokes parameters Q of the
back-scattering photons:
\begin{eqnarray}
Q_{bs} \propto \frac{\pi}{2}\sin\theta_0\sin^2
i(1-\sin^2\theta_0)-\frac{2}{3}\sin2i(1-\cos^3\theta_0)
\end{eqnarray}
This equation tells that the back-scattered light has always
$Q_{bs} > 0$ (i.e. polarization parallel to the symmetric axis)
when models A-12$^{\rm o}$ and B-33$^{\rm o}$-20$^{\rm o}$ are viewed
at the BALR directions ($90$$^{\rm o}$$-\theta_0\leq i\leq 90$$^{\rm o}$).
However, the polarization can be either parallel, i.e., $Q_{bs} >0$ at
$i=60$$^{\rm o}$, or perpendicular to the symmetric axis, i.e., $Q_{bs} < 0$
at $i=45$$^{\rm o}$ for model B-45$^{\rm o}$-30$^{\rm o}$.
Furthermore the selective leakage is working only at $v>-v_{\varphi,0}$
(see \S3.2.1) while most resonantly scattered photons will be at small
velocities (see \S3.1.1), therefore, back-scattering can dominate the
polarized light in the blue part of trough. This explains why the blue
PA swing appears only in the model B-45$^{\rm o}$-30$^{\rm o}$ viewed
at $i=45^{\rm o}$ for all cases considered. Our interpretation is
verified by our simulation and presented in Fig \ref{b45se}.

The red PA swing (Figs. \ref{b33se} and \ref{b45se}) is caused by
the same reason as the sub-trough for model with SESR. We will
discuss this in details in the next section.

\section{Doublet Transitions}

In this section we will simulate the resonant scattering for doublets,
and compare the results with those for singlets. As an example, we
consider the two transitions of CIV: $J=1/2\rightarrow1/2$
($T_{1/2}$, $\lambda=$1550\AA) and $J=1/2\rightarrow3/2$
($T_{3/2}$, $\lambda$=1548\AA). The cross section of the former
transition is 1/2 of the latter. According {\em Eq.} 4, the
scattered light of $T_{1/2}$ transition is unpolarized regardless
the polarization of the incident light.

\subsection{Small Electron Scattering Region}

\subsubsection{Model A}
\begin{figure}
\plotone{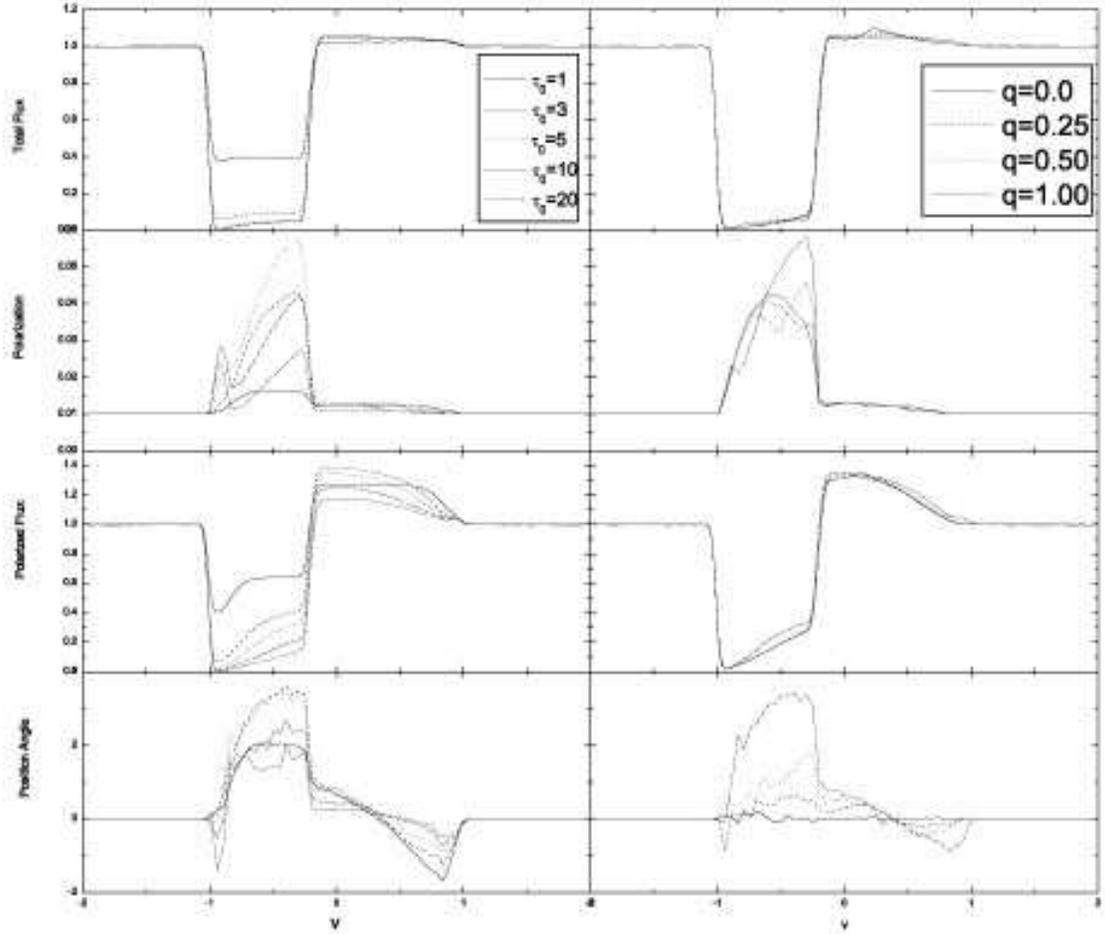} \caption{Total flux, polarization degree,
polarized flux and the PA of polarization for scattering of
doublet transition. The model adopt is A-12$^{\rm o}$ and a SESR.
In the left panel, $q=1.0$ is constant and different $\tau_0$ is
marked. And in the right panel, $\tau_0=5$ and different $q$ is
marked.
 Here all the four quantities are the
average value over viewing angle for the BAL QSOs.}\label{a12dne}
\end{figure}

We first make simulations for model A4-12$^{\rm o}$ with total
optical depths of the doublets $\tau_0$=1, 3, 5, 10, 20. The
output PA, polarization degree and polarized flux are showed in
Fig. \ref{a12dne}. In many aspects, the scattering of doublets
produce similar characteristics as the scattering of singlets but at
reduced polarization degree: PA rotation and asymmetric profile of PA
rotation at moderate and large optical depths, larger
polarization in the trough than in the continuum, a shallower
trough in the polarized flux and the excess of the polarized flux
across the emission line position. However, there are apparently
differences too. First, the polarization degree is greatly reduced,
with a maximum of only $6.0\%$ at $\tau_0=5$. Second, the PA
rotation decreases with increasing optical depth for $\tau_0>5$
though the maximum ($\sim 5$$^{\rm o}$ for model A4-12$^{\rm o}$)
is similar to that for singlet.

The differences can be understood as follows. For an accelerating
outflow, the continuum photons encounter the scattering surface of
$T_{3/2}$ transition first then the $T_{1/2}$ transition surface,
which will erase the polarization produced by $T_{3/2}$ surface
(Blandford et al. 2002), especially, at large optical depth. As a
result, the final polarization degree is greatly reduced, and is
essentially determined by the optical depth of $T_{1/2}$ surface
when $\tau_0$ is moderate or large. As the optical depth
increases, the erasion of polarization by $T_{1/2}$ scattering
surface is more effective in the inner region, where large PA
rotation of the scattered light from $T_{3/2}$ surface is
produced. This leads to a decrease of PA rotation with the
increase of optical depth.

High polarization in CIV and NV troughs observed in some BAL QSOs
requires rather small $\tau(v)$ for the transition $T_{1/2}$, at least
in a significant part of the flow if the polarization is
attributed to the resonant scattering. For the models described in \S
2,  if $\beta$ is substantially larger than 0.5, the optical depth
$\tau$ in radial direction  will drop rapidly outward ({\em Eq.
3}), and eventually becomes quite moderate at large radii, to the
transition $T_{1/2}$ even the inner region is optically thick. The
Monte-Carlo simulations have been carried out for $\beta=$0.75,
1.5, 2.0 and $\tau_0=3$. We find that the maximum polarization
increases with increasing $\beta$, and reaches $\sim$ 9\% for
$\beta=2$. We also made simulations for different $\theta_0$, and
find that the polarization degree decreases slightly and PA rotation
increases with increasing $\theta_0$ in the optically thick cases.
The reason is that the flux in the trough is already dominated by
the scattered light, of which the polarization decreases slightly with
increasing $\theta_0$. These results illustrate that the
polarization depends much more strongly on the velocity structure
than on $\theta_0$ for a moderate or large optical depth.

LB97 considered an equatorial outflow that is accelerated on
hyperboloid surfaces, and obtained a higher maximal polarization
in the trough, 50\% for singlets and 15\% for doublets (10\% when
rotation is included). The main cause for such difference is that
they adopted a different prescription of the flow that moves in
poloidal (and azimuthal) direction as such the velocity along
the line of sight is non-monotonic. They also got a small polarization
($\simeq$ 10\% for singlets) for velocity law $v_r\propto r$.

\subsubsection{Model B}

\begin{figure}
\epsscale{1}\plotone{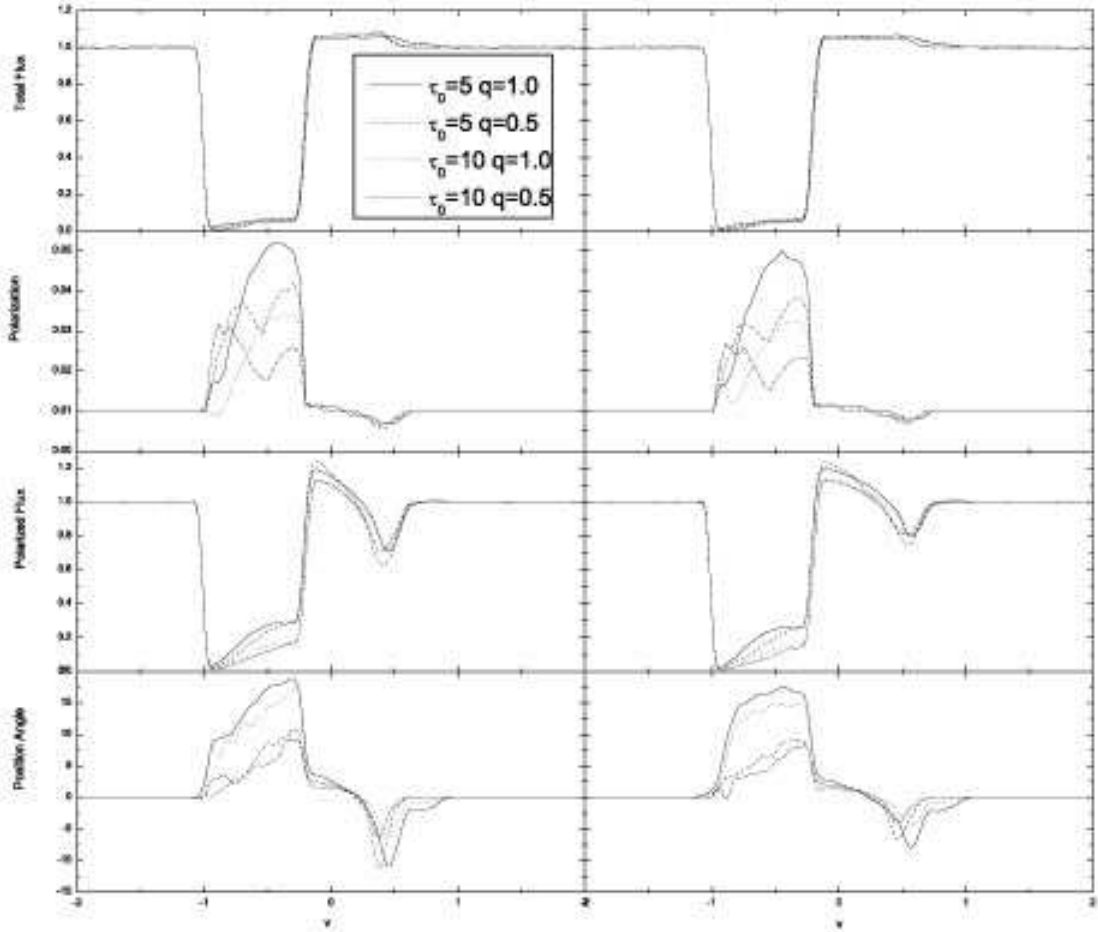} \caption{The total flux,
polarization degree, polarized flux and PA for doublets . The
model adopted is B-33$^{\rm o}$-20$^{\rm o}$ and a SESR. The left
panel is viewing from $i=57^{\rm o}$ and the right one is viewing
from $i=70^{\rm o}$. $\tau_0$ and $q$ are marked in the figure.
}\label{b33dne}
\end{figure}

\begin{figure}
\plotone{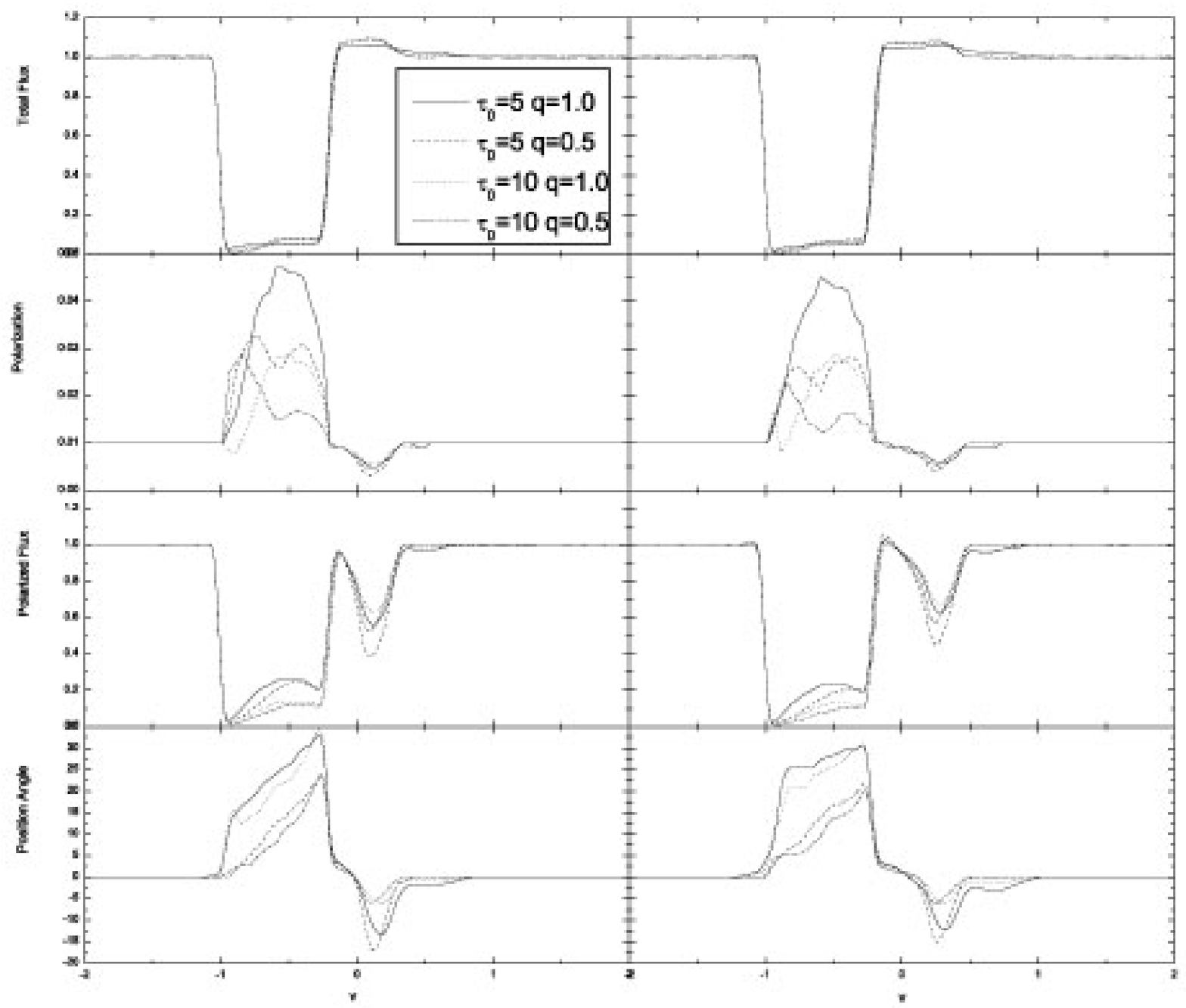} \caption{ Similar to Fig. \ref{b33dne} but for
model B-45$^{\rm o}$-30$^{\rm o}$, viewing from $i=45^{\rm
o}$(left panel) and $i=60^{\rm o}$(right panel). }\label{b45dne}
\end{figure}

The numerical results for models B-33$^{\rm o}$-20$^{\rm o}$ and
B-45$^{\rm o}$-30$^{\rm o}$ are summarized in Table 1 and shown in
Fig. \ref{b33dne} and Fig. \ref{b45dne}.

\begin{figure}
\plotone{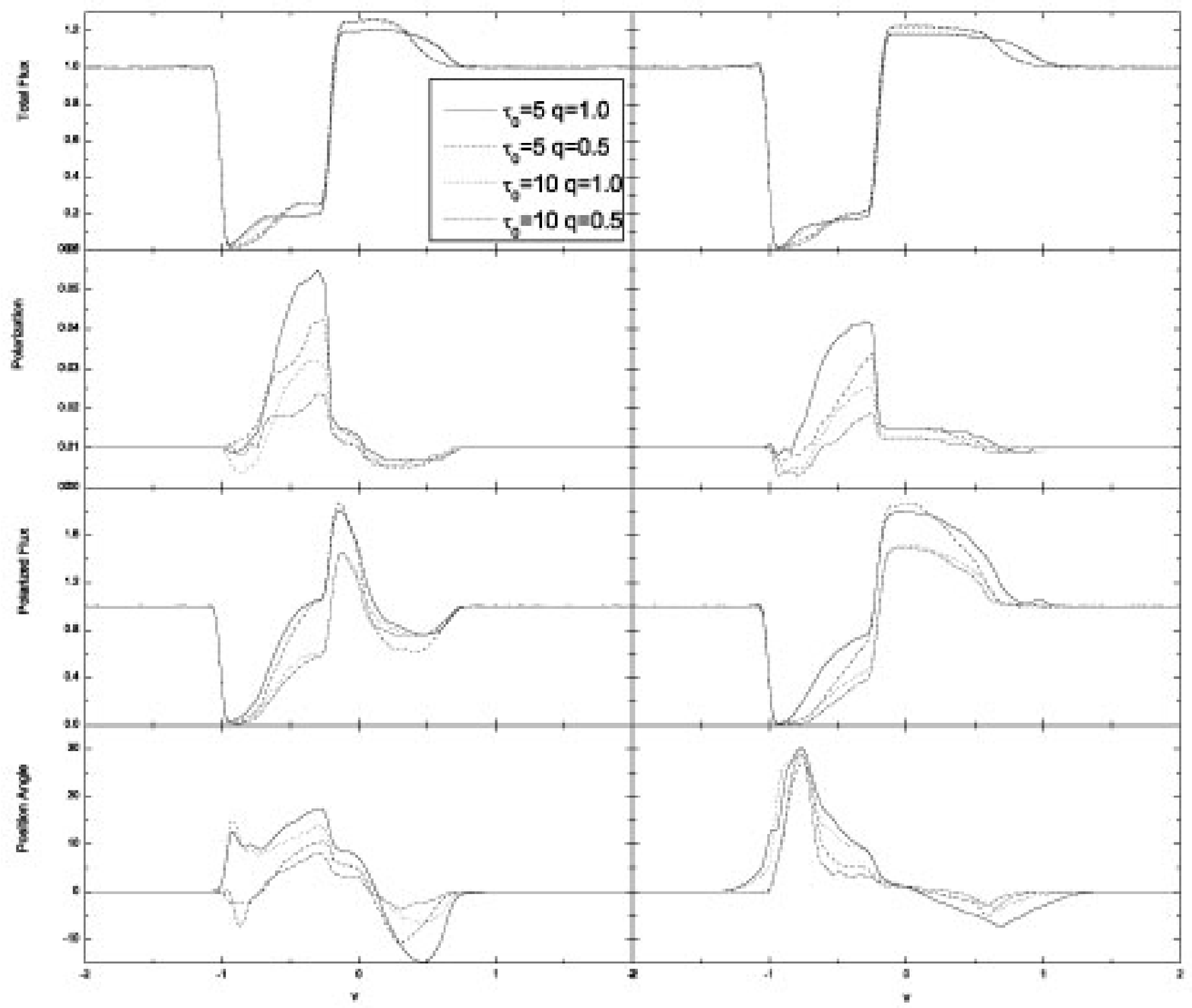} \caption{ Similar to Fig. \ref{b33dne} but for
model A-45$^{\rm o}$, viewing from $i=45^{\rm o}$(left panel) and
$i=90^{\rm o}$(right panel). }\label{a45dne}
\end{figure}

Similar to singlet models, an additional absorption trough
appears to the red side of the trough in the polarized flux. This
sub-trough is especially prominent for large $\theta_0$'s and at
small inclinations. As $\theta_0$ decreases and viewing angle
increases, it becomes shallower and moves to the red side, and an
emission-line like feature appears around $v=0$. PA rotation is
also much larger than those in model A-12$^{\rm o}$ and can reach
as high as 30$^{\rm o}$  for $\theta_0=45$$^{\rm o}$ when
$\tau_0=5$. From Fig. \ref{a45dne}, it is clear the PA rotation
across the profile is sensitive to the inclinations as well. PA
rotation peaks at small velocities at small inclinations,
but the peak shifts to the blue part of the trough at large
inclinations. However, large PA rotation and the appearance of the
red sub-trough in the polarized light is a characteristic of
all models with large $\theta_0$, rather than of type B-models only
(Figs. \ref{a45dne} and \ref{sub-q0}). But the shape of sub-trough is
different for type-A and for type-B models: broad for type-A models and
narrow for type-B models. A comparison of polarized line profile for
different models is given in Fig. \ref{sub-q0}. We find that when
$\theta_0$ increases the sub-trough becomes broader and deeper,
and shifts to the blue-side. As $\theta_1$ decreases it gets broader and
shallower and shifts to the red side, and an emission-line like
feature appears due to the scattered photons from the low
latitude. Only models with large $\theta_1$ can produce an
apparent sub-trough, which is much narrower than the primary
trough.

The red sub-trough in the polarized flux actually is not due to
absorption, but result from the cancellation of the continuum
polarization by the scattered polarized line flux as shown in
Appendix A. In the Appendix, we demonstrate that the necessary
condition for the appearance of the sub-trough is $\theta_0\gtrsim 25^{\rm o}$
regardless the rotation velocity of the flow. Hence
the appearance of this feature sets a lower limit on $\theta_0$.
We also compare the numerical results for models A0-12$^{\rm o}$,
A0-20$^{\rm o}$,
A0-25$^{\rm o}$, A0-30$^{\rm o}$ \& A0-45$^{\rm o}$ with models
B0-20$^{\rm o}$-10$^{\rm o}$, B0-25$^{\rm o}$-10$^{\rm o}$, B0-30$^{\rm o}$-10$^{\rm o}$,
B0-33$^{\rm o}$-20$^{\rm o}$ \& B0-45$^{\rm o}$-30$^{\rm o}$, and find
a sub-trough in the polarized flux only in models with $\theta_0\geq25^{\rm o}$
(Fig. \ref{sub-q0}). Numerical results in Fig. \ref{sub-q0} and
Fig. \ref{com-pa90} (PA$_c$=0$^{\rm o}$) also confirm that even if
$q=0$ the sub-trough also appears. The sub-trough in the A0 and B0
models appears to be deeper than that of the A4 and B4. But the
depth of the sub-trough depends on optical depth and is inversely
proportional to the continuum polarization.

The numerical results are summarized in the Table 2 for $\tau_0=5$
 and four different $q$'s, and some of them are
plotted for comparison (Fig. \ref{b33dne} and \ref{b45dne}). PA
rotation increases with increasing $q$, as for singlets, and depends
strongly on $\theta_0$. The maximum polarization in trough, about
$4\%\sim6\%$, decreases slightly with increasing $\theta_0$ and
increases with increasing $q$, same as what we noticed in
previous section.

\begin{table}[h]
\begin{center}
\caption{Maximum PA rotation(PA$_m$) and polarization($P_m$) for
different models} \label{lab}
\begin{tabular}{ccccc}
  \hline\hline
  $4\times q$ & 1 & 2 & 3 & 4 \\
  \hline
  & & PA$_m$ & &\\
  \hline
  A-12$^{\rm o}$ & -- & 2$^{\rm o}$$\sim$3$^{\rm o}$ & -- & 3$^{\rm o}$$\sim$5$^{\rm o}$ \\
  A-30$^{\rm o}$ & 2$^{\rm o}$$\sim$8$^{\rm o}$ & 5$^{\rm o}$$\sim$12$^{\rm o}$ & 6$^{\rm o}$$\sim$14$^{\rm o}$ & 9$^{\rm o}$$\sim$16$^{\rm o}$ \\
  A-45$^{\rm o}$ & 3$^{\rm o}$$\sim$24$^{\rm o}$ & 7$^{\rm o}$$\sim$28$^{\rm o}$ & 12$^{\rm o}$$\sim$32$^{\rm o}$ & 16$^{\rm o}$$\sim$31$^{\rm o}$ \\
  B-33$^{\rm o}$-20$^{\rm o}$ & 4$^{\rm o}$$\sim$6$^{\rm o}$ & 8$^{\rm o}$$\sim$11$^{\rm o}$ & 13$^{\rm o}$$\sim$16$^{\rm o}$ & 15$^{\rm o}$$\sim$19$^{\rm o}$ \\
  B-45$^{\rm o}$-30$^{\rm o}$ & 10$^{\rm o}$$\sim$14$^{\rm o}$ & 21$^{\rm o}$$\sim$26$^{\rm o}$ & 28$^{\rm o}$$\sim$34$^{\rm o}$ & 31$^{\rm o}$$\sim$35$^{\rm o}$ \\
  \hline
  & & $P_m$ & & \\
  \hline
  A-12$^{\rm o}$ & 4.5\% & 4.9\% & 5.5\% & 6.0\% \\
  A-30$^{\rm o}$ & 4.3\% & 4.8\% & 5.4\% & 5.9\% \\
  A-45$^{\rm o}$ & 3.8\% & 4.2\% & 4.8\% & 5.5\% \\
  B-33$^{\rm o}$-20$^{\rm o}$ & 4.2\% & 4.2\% & 4.8\% & 5.2\% \\
  B-45$^{\rm o}$-30$^{\rm o}$ & 3.9\% & 3.3\% & 4.0\% & 4.8\% \\
  \hline
\end{tabular}
\end{center}
\end{table}

\subsection{Large Electron-Scattering Region}

\begin{figure}
\plotone{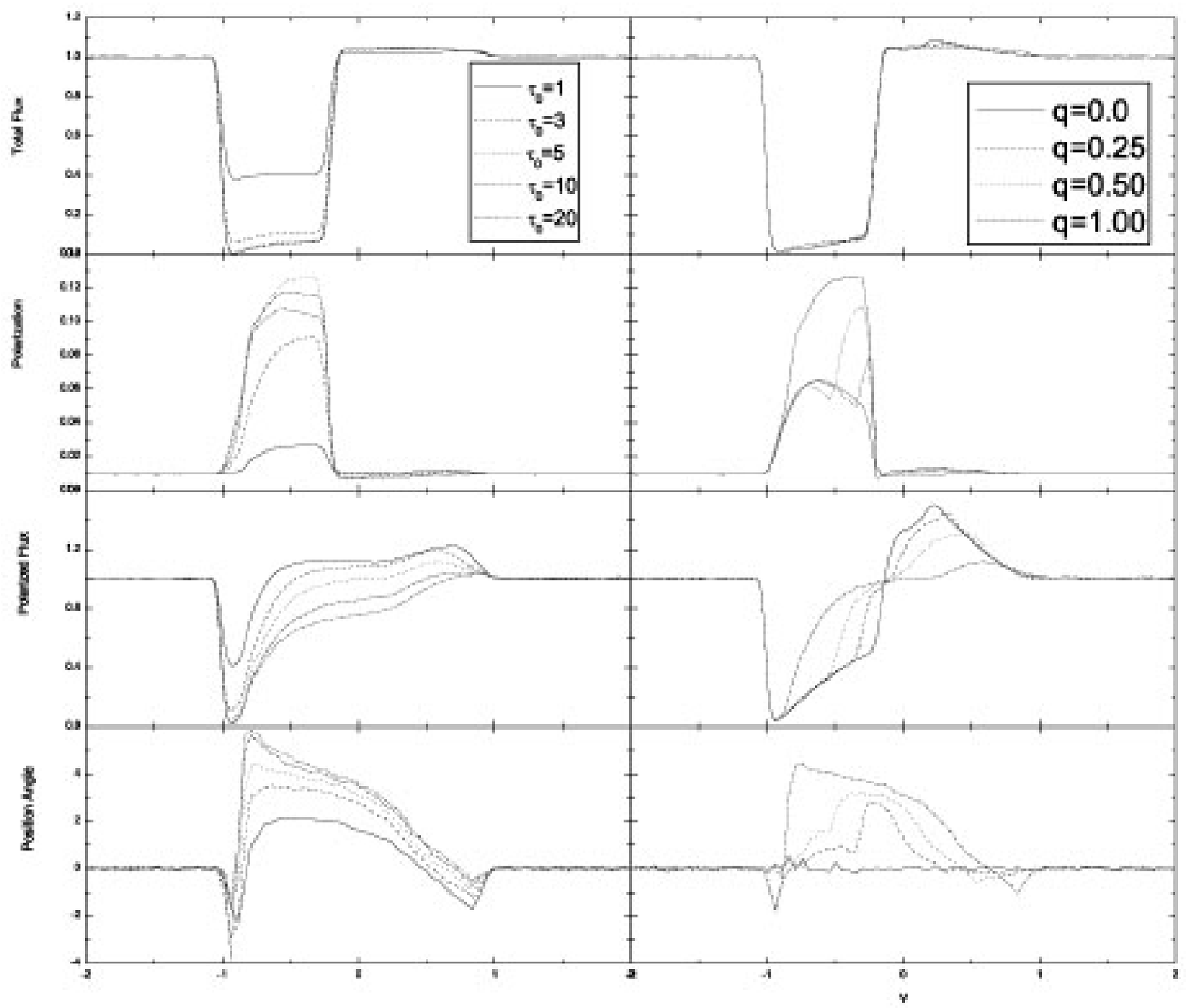} \caption{ Similar to Fig. \ref{a12dne} but for
LESR. }\label{a12de}
\end{figure}

\begin{figure}
\plotone{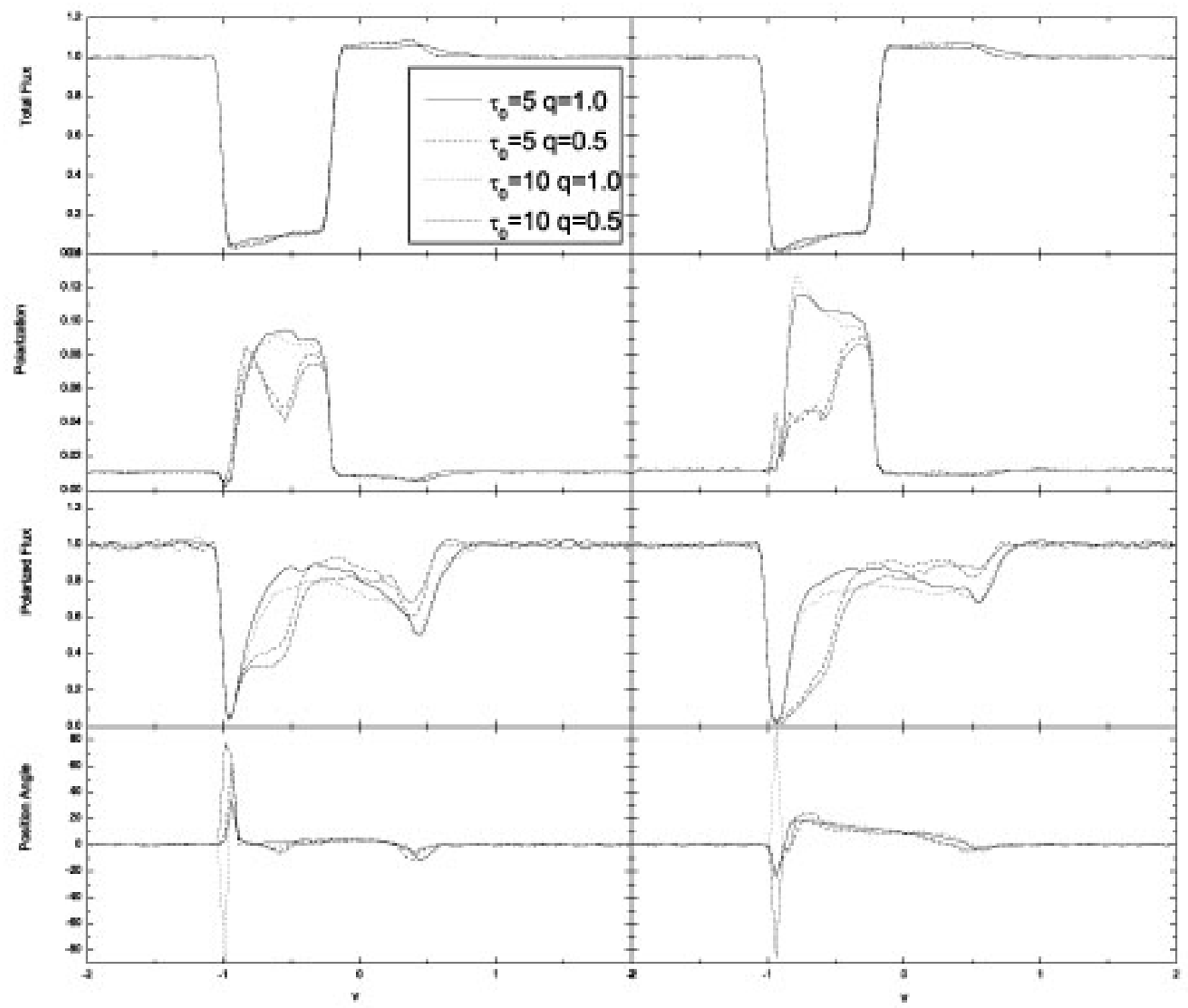}
 \caption{
Similar to Fig. \ref{b33dne} but for LESR.
}\label{b33de}
\end{figure}

\begin{figure}
\plotone{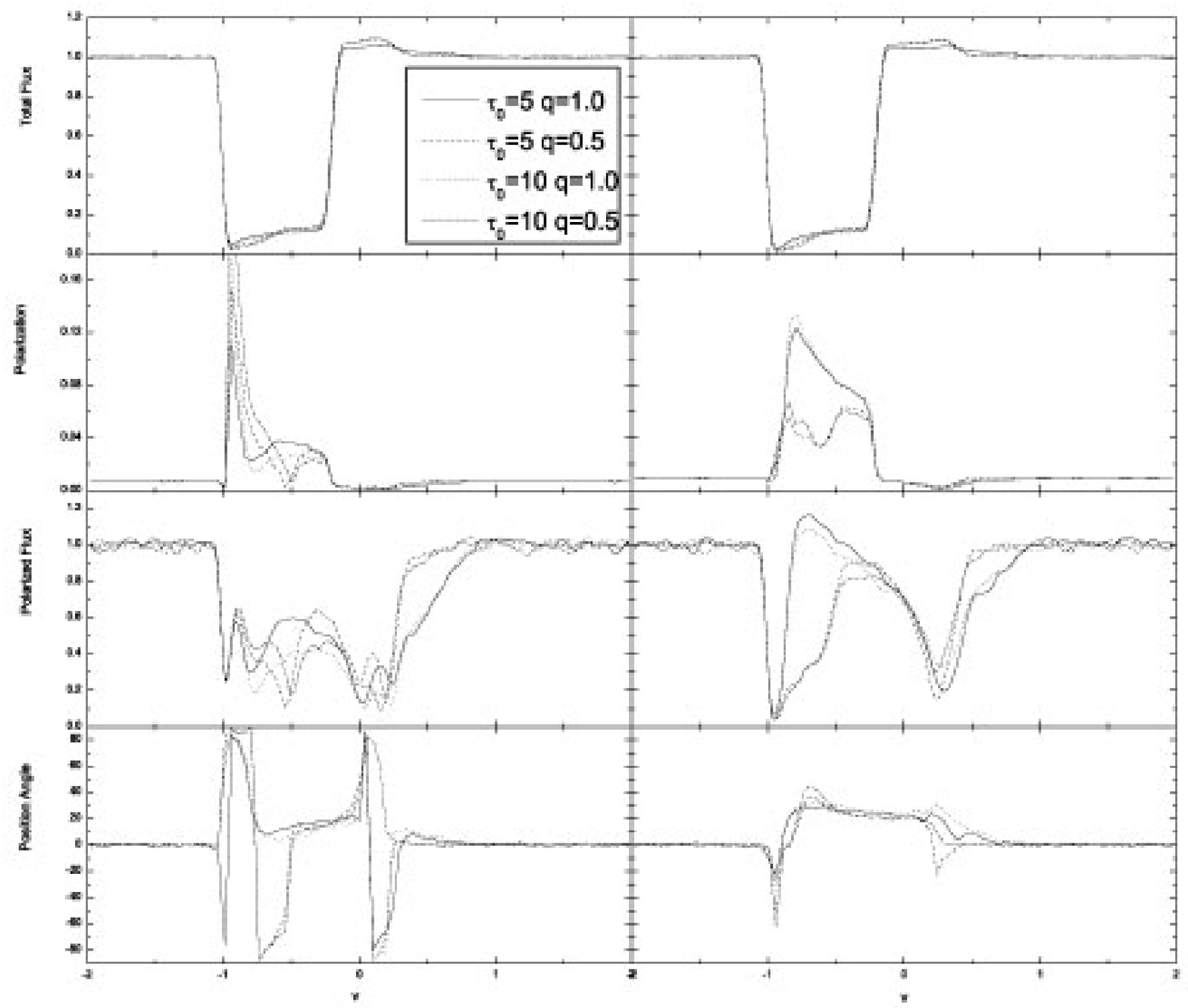}
 \caption{
Similar to Fig. \ref{b45dne} but for LESR.
}\label{b45de}
\end{figure}

\begin{figure}
\epsscale{0.45}\plotone{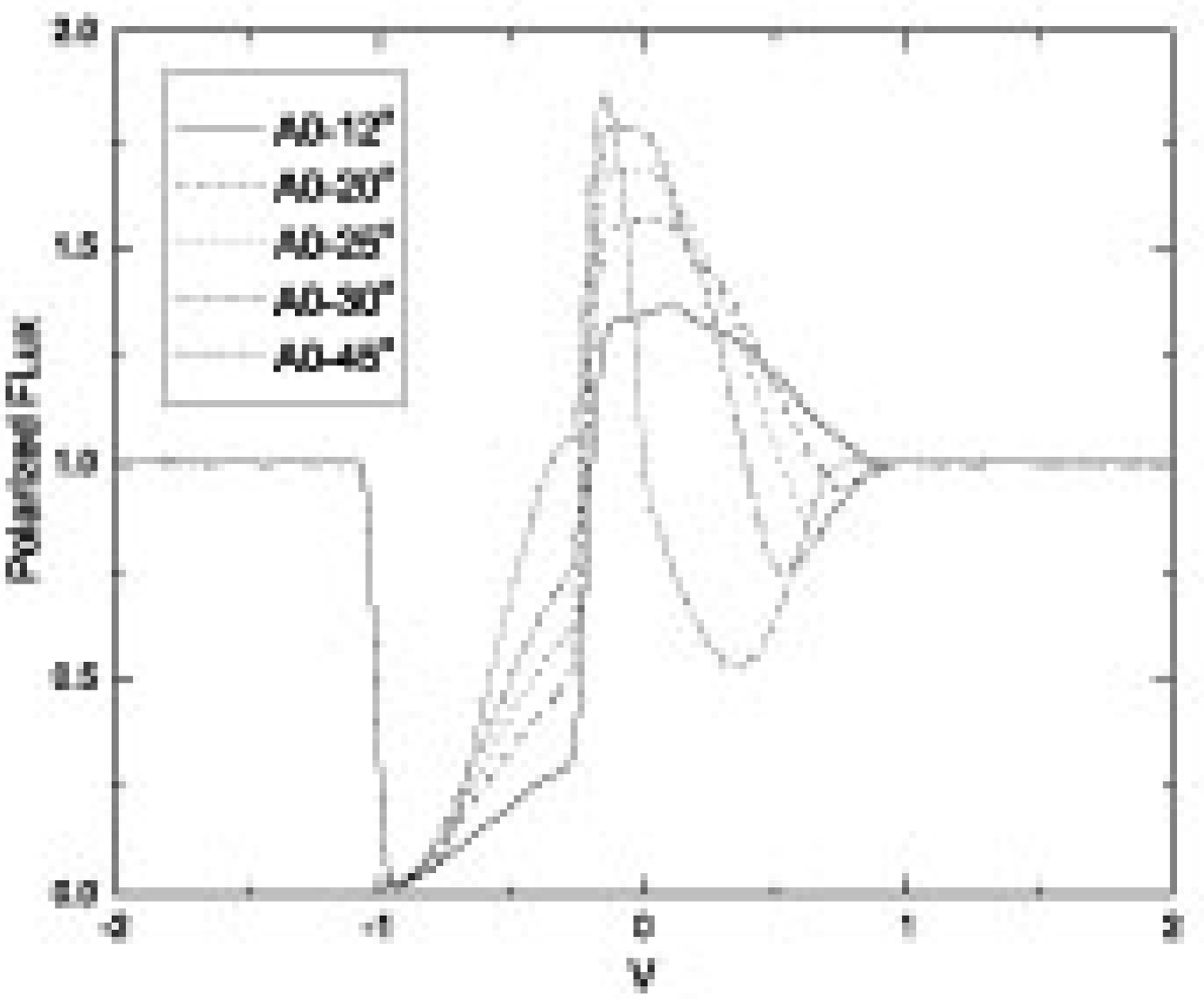}\plotone{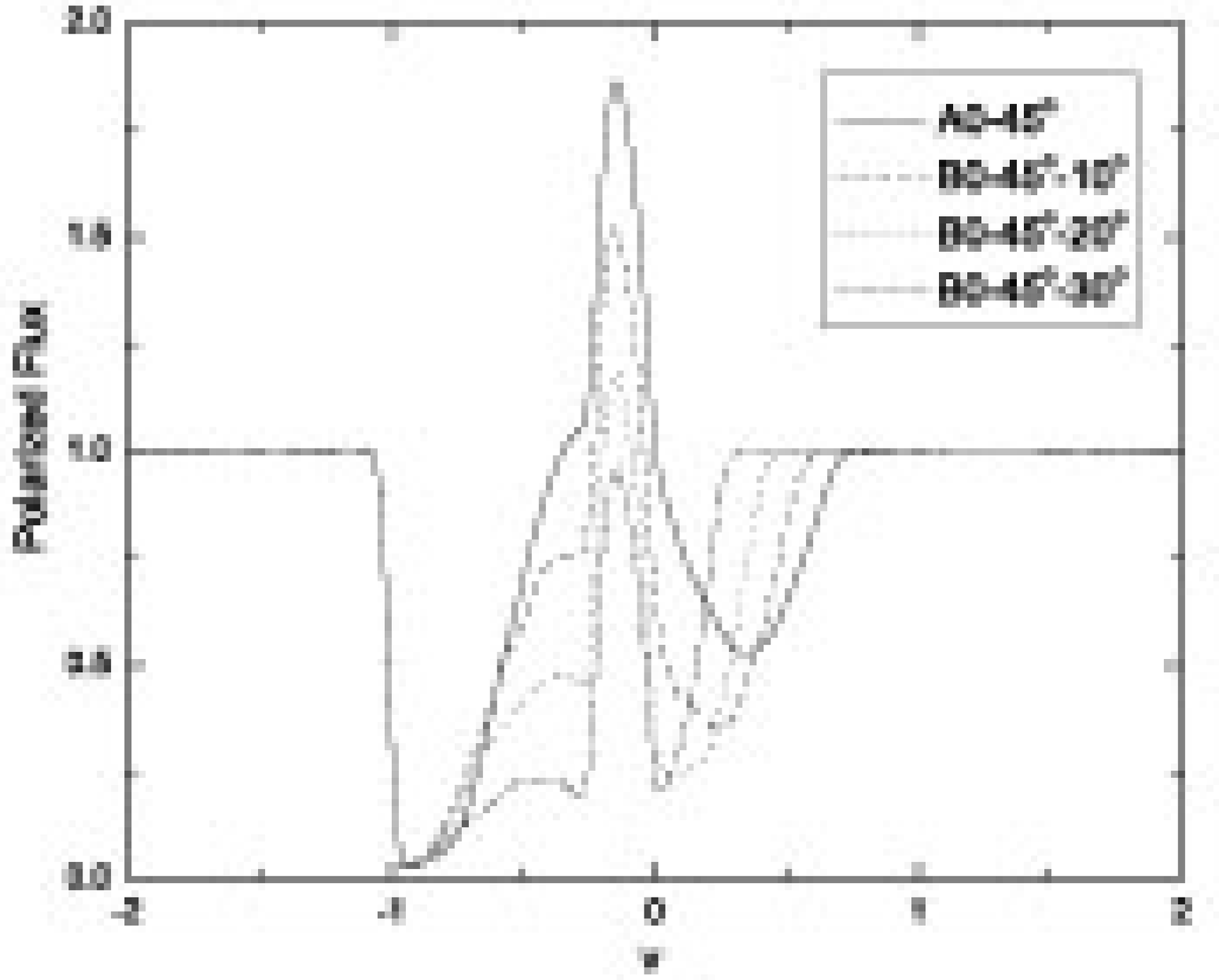}
\plotone{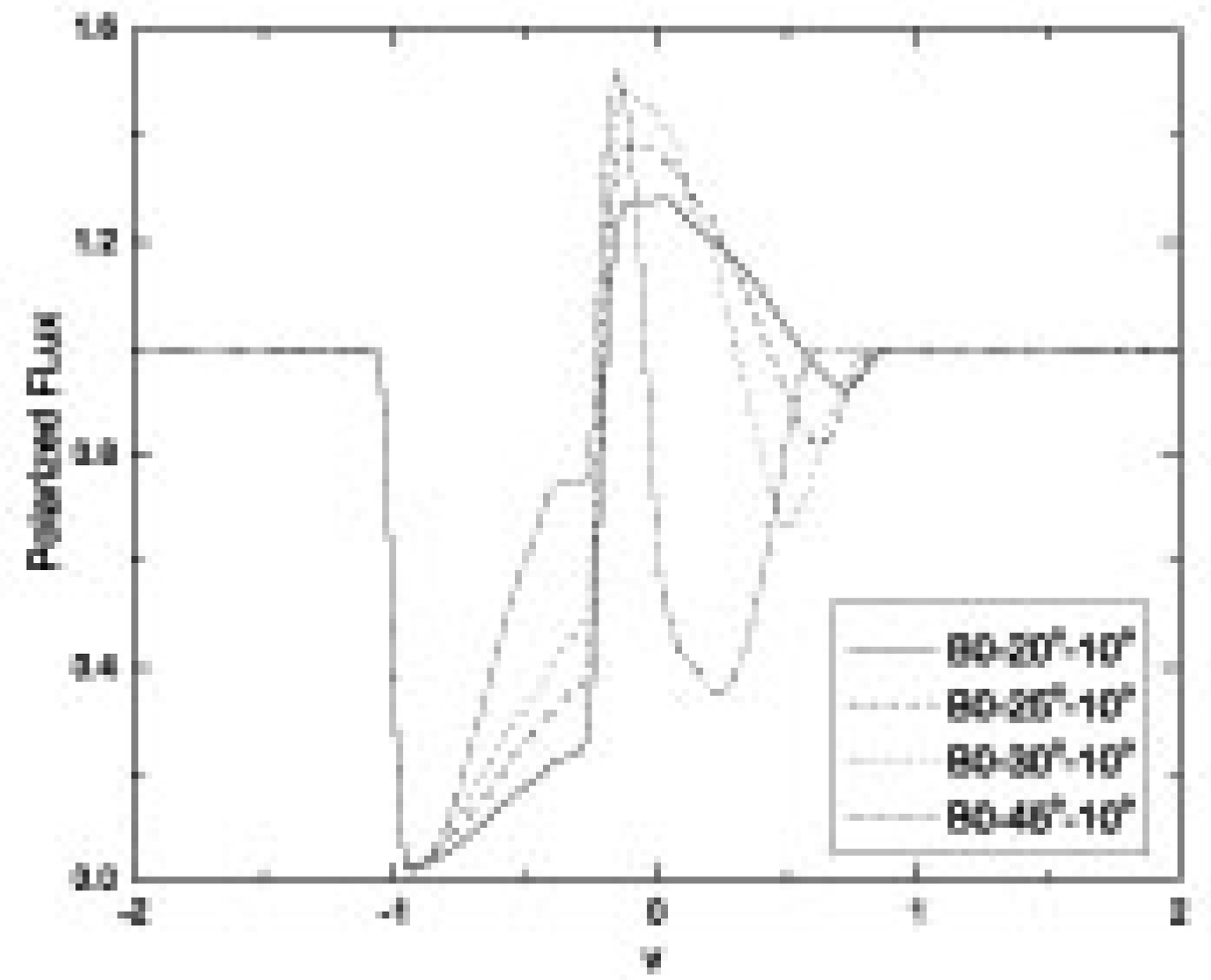} \caption{Polarized flux from scattering of
doublets for different $\theta_0$ and $\theta_1$ viewed from
$i=90$$^{\rm o}$$-\theta_0$. $\tau_0=5$ is assumed in all models
and with a SESR on the base of the outflow.}\label{sub-q0}
\end{figure}

\begin{figure}
\epsscale{0.45} \plotone{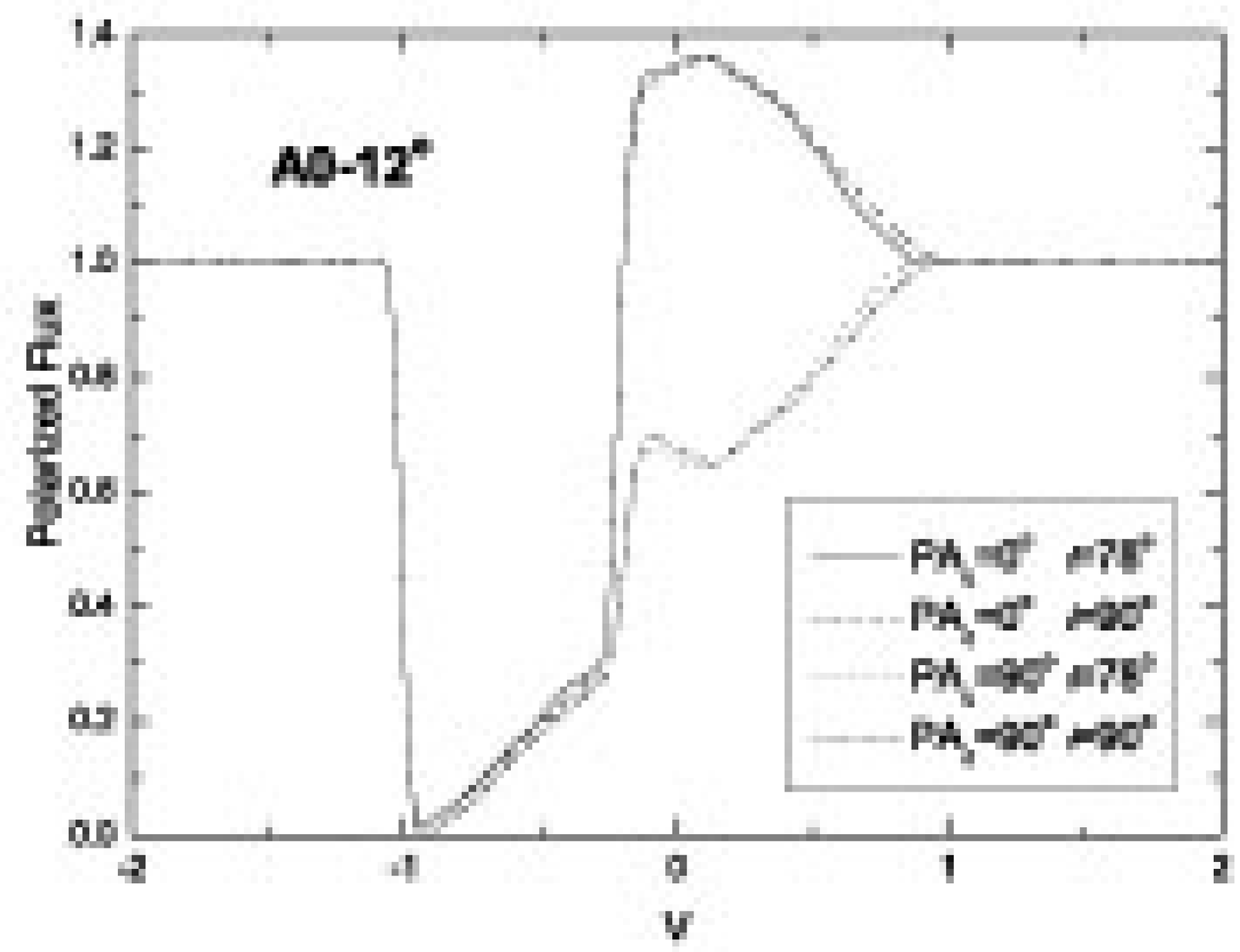}\plotone{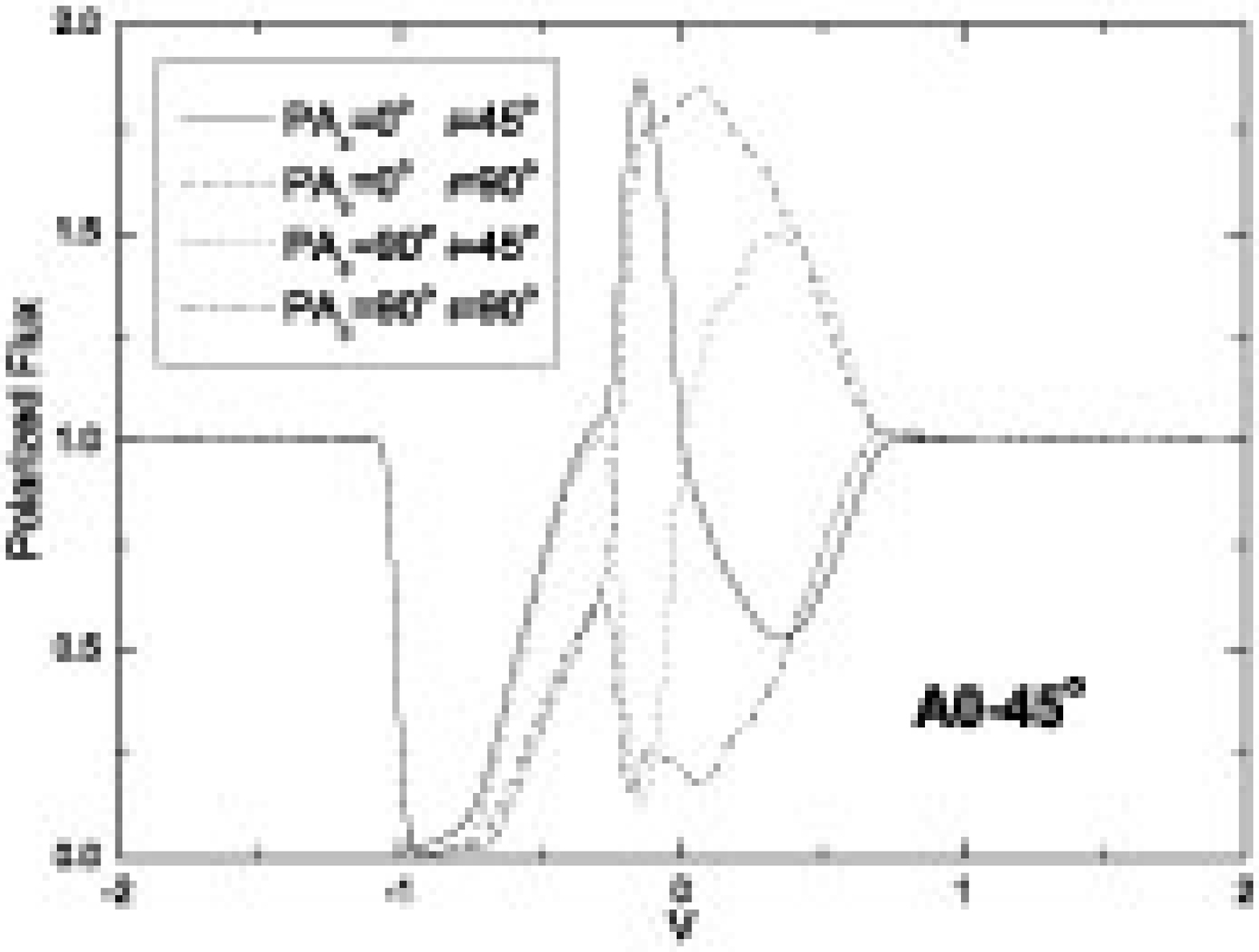}
\plotone{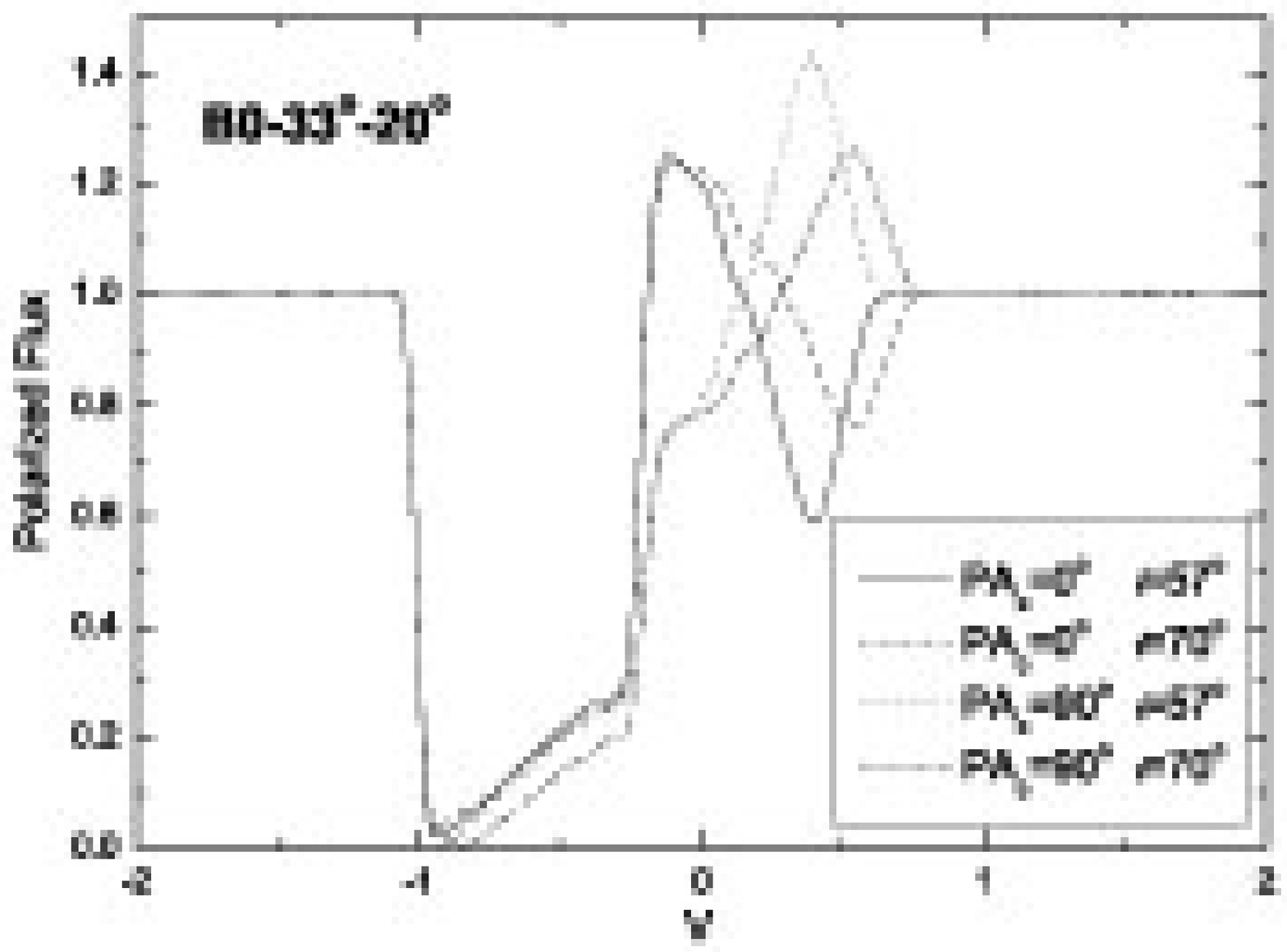}\plotone{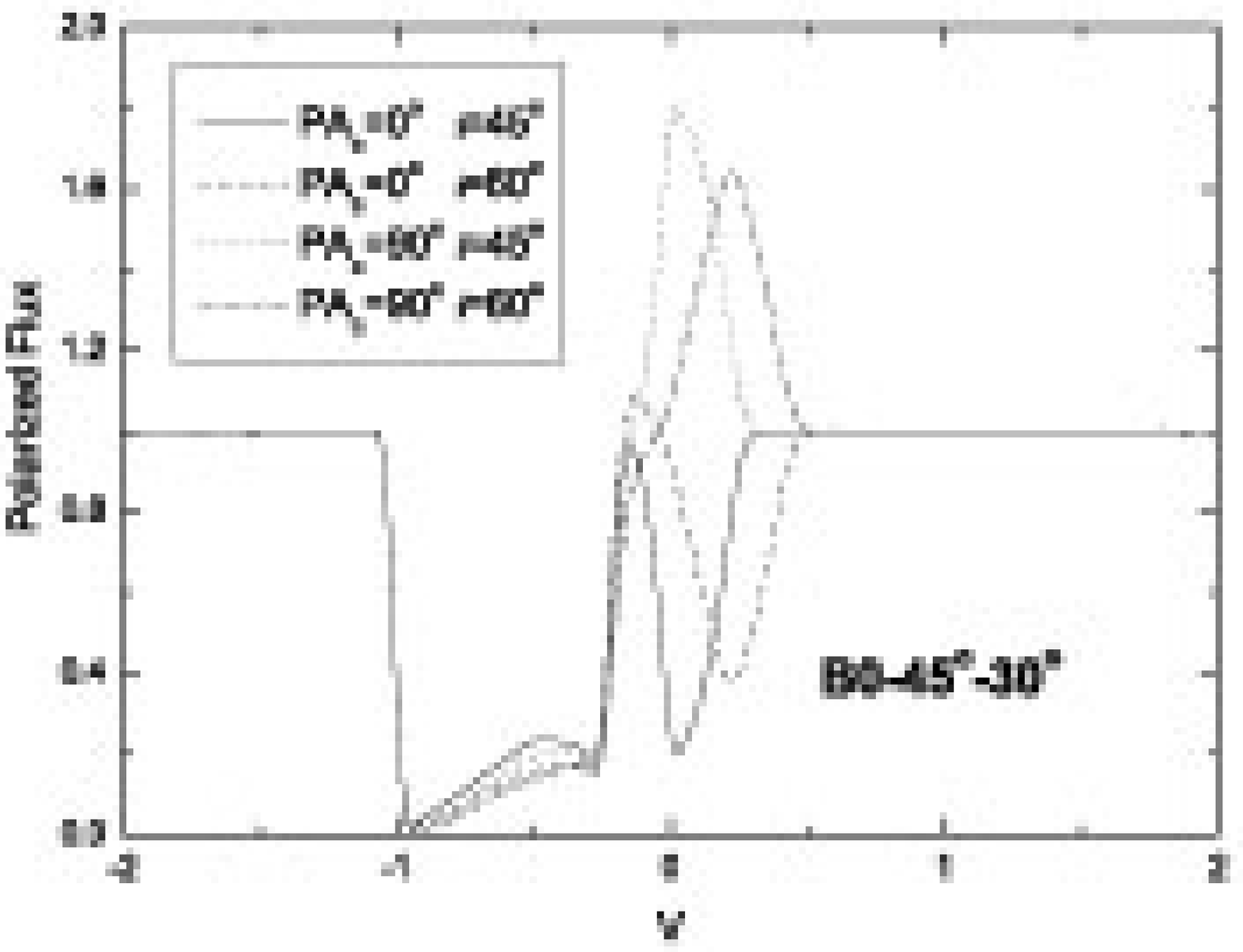} \caption{ Polarized flux from
scattering of doublet for two incident continuum polarization,
PA$_c=0$$^{\rm o}$ and PA$_c=90$$^{\rm o}$, and for model
A0-12$^{\rm o}$, A0-45$^{\rm o}$, B0-33$^{\rm o}$-20$^{\rm o}$ and
B0-45$^{\rm o}$-30$^{\rm o}$. $\tau_0=5$ is assumed.
}\label{com-pa90}
\end{figure}

For doublets, we carry out simulations also for LESR models.
The results are shown in Fig. \ref{a12de} for type-A models, Fig.
\ref{b33de} and \ref{b45de} for type-B models. Due to the
erasion of $T_{1/2}$ transition, the electron scattered light dominates
the polarized flux in the trough when $\tau_0$ is large. As a
result, the polarized flux and PA rotation are more similar to
WCRS models than singlet models, whereas the summed polarization
of the resonantly and electron scattered photons is lower for
doublets than for singlets.

As demonstrated in the Appendix A, when $\theta_0$ is large enough
the Stokes parameter $Q_l$ for the resonantly scattered light at
positive $v$ 
becomes negative, i.e., PA$_l$ is about 90$^{\rm
o}$. For SESR models with polarization of the incident continuum
$\sim$1\%, $|Q_l|$ is generally smaller than $|Q_c|$ in the entire
velocity range, as such the final polarization has the same direction
as the continuum (PA$_c\sim 0^{\rm o}$) and only a sub-trough
appears to the red side of the line center. However, in LESR models, part of
red-shifted scattered continuum is absorbed so that $|Q_l|$ in the
bottom of the sub-trough may be larger than $|Q_c|$. This leads to
PA$_l=90^{\rm o}$ and a local excess of the polarized flux at
these velocities (Fig. \ref{b33se}, \ref{b45se} and \ref{b45de}).
If $|Q_l|>2|Q_c|$ the peak is higher than the polarized flux of
the continuum (see Eq. \ref{app1} in Appendix A). We find such cases
only for singlet models (Fig. \ref{b45se}). This feature appears
only in certain geometry models and for certain ranges of $\tau_0$
and continuum polarization degree.

To summarize, resonant scattering of doublets usually produces
much lower polarization at large optical depth for
accelerating flows. The polarization degree is sensitive to the velocity
law. A slowly accelerating flow (large $\beta$) or non-monotonic
velocity in radial directions will produce higher polarizations.
Models with LESR can easily produce high polarization due to back-scattering
and leakage of the electron scattered continuum. Large PA rotation can
be reproduced only by a rotating outflow with a large $\theta_0$ for
either SESR or LESR. Models with large subtending angle $\theta_0$ can
produce a sub-trough to the red side of the primary trough in the
polarized flux regardless the rotation velocity. A jump appears in
PA rotation and polarized flux across the starting velocity of the
outflow in SESR models, but it disappears in LESR models.
The absorption to the polarized flux can extend to $v=v_{\varphi,0}$
at the red side of the line center for LESR models at large $\tau_0$.

\section{Comparison with Observations}

\subsection{Polarization Degree}
The observed polarization degree in the BAL line troughs vary from
non-detection to up to 20\% for strong doublets such as CIV and NV
(O99). While the polarization degree at a few percent level can be
reproduced by most models considered, polarization degree higher
than 10\% for doublets detected in a few objects puts severe
constraints on the models. Below we will discuss three possible
models that may reproduce such high polarization: a decelerating
flow, a flow with a velocity law similar to that of LB97, and a
large electron-scattering region.

Although it is generally believed that the outflow is
accelerating, there is no compelling evidence against a possible
decelerating region in the outer part of the flow (Voit, Weymann,
\& Korista 1993). Such non-monotonic velocity is observed in
outflows of a handful Seyfert galaxies on sub-kpc scales (e.g, Ruiz
et al. 2001). We notice that recent X-ray observations have
revealed X-ray BALs which require massive outflow with higher
velocities at radii smaller than UV BALR (Chartas et al. 2002;
2003). This also supports the existence of a decelerating region
in the outflow. High polarization will be produced because the
continuum photons are scattered by the surface of transition
$T_{3/2}$ after they encountered $T_{1/2}$ so that the polarized
light produced by scattering of the transition $T_{3/2}$ can reach
us freely. With respect to scattering by singlet, there are two
differences. First, the effective optical depth used to produce
the polarized flux is contributed by the transition $T_{3/2}$
only, substantially lower than the total optical depth. Second,
the photons reaching $T_{3/2}$ surface is a combination of
transmitted light (through $T_{1/2}$ surface) and the scattered
light, which appears much more isotropic to the scattering ions.

LB97 showed that large polarization can also be produced in a
monotonic accelerating flow which follows a different prescription
of velocity-law. As mentioned in previous section the velocity
along the line of sight is also non-monotonic in their model. In
previous section, we find models with a LESR can also produce high
polarization degree because electron-scattered photons can fill
the troughs and increases the polarization.

Ly$\alpha$ is actually a special doublet, in which fine structure
splitting is very narrow, about 1.3~km/s, comparable to the
thermal velocity. In that case, the polarization can be much
higher because there is a good chance that polarized scattered
photons (from $T_{3/2}$) escapes before encountering $T_{1/2}$
scattering. However, some of these escaped photons will be
scattered again by NV at ISV surface with a velocity difference of
-5,900~km~s$^{-1}$ from the current Ly$\alpha$ scattering surface,
so that their polarization will be erased again. At high altitude,
the back-scattered light usually do not meet the NV scattering
surface, and produce an excess in polarized flux in the red side
of the Ly$\alpha$ emission line. The polarization in the overlap
region of NV and L$\alpha$ is not as high as expected for the
scattering of Ly$\alpha$ alone, but in general be higher than the
pure NV scattering because a fraction of scattered Ly$\alpha$
photons comes directly. Since scattering by NV occurs after
Ly$\alpha$ scattering, it is proper to consider the NV scattering
with the incident light including scattered L$\alpha$ photons in
the overlapped wavelengths. At flow velocity greater than
$v_0$+5,900 km~s$^{-1}$, NV ions see a polarized continuum at
resonant frequency of the ion with a finite size formed through
Ly$\alpha$ scattering. At lower velocities, both direct continuum
and the scattered Ly$\alpha$ continuum may contribute to the
incident continuum being scattered. The geometry of the scattered
Ly$\alpha$ photon at the resonant frequency of NV is complex. As
such some characteristics of the LSER models may be retained in
the NV scattered light regardless whether the electron scattering
region is extended or not. The polarization will be enhanced in
this case.
The apparent excess of polarized flux to the red side of Ly$\alpha$
emission line observed in some BAL QSOs (O99) is likely caused by
these effects. We leave a detailed analysis of Ly$\alpha$ scattering
to a future work.


\subsection{Position Angle Rotation}

The PA rotation in BAL trough appears common, but accurate
measurements are still rare. To produce the PA rotation in an
axisymmetric outflow model, the flow must carry a substantial
angular moment. Besides, the sign of the PA rotation is an
indicator of the angular moment direction. Our simulations show
that models with and without LESR can both produce PA rotation.
The main difference between the two type models is that a jump in
both PA and polarized flux appears at the starting velocity of the
absorption trough in all models without LESR but not models with
LESR. However we can not put the quantitative constraint on the
rotation from current observations because of the poor S/N.

O99 detected PA rotations as large as $20^{\rm o}$$\sim 30$$^{\rm
o}$ in BAL troughs of several BAL QSOs. Numerical calculations
suggest that flows should extend to at least $\theta_0 \simeq
30$$^{\rm o}$ in order to account for PA rotation larger than
15$^{\rm o}$ (see Table 2). The PA of polarization varies across
the line profile in simulation, and its profile depends also strongly
on the inclination and velocity distribution, but very weakly
on the optical depth for the large $\tau_0$ models
considered (Fig. \ref{b33dne} to Fig. \ref{b45de}). For a wide
range of models, the maximum PA rotation occurs at a small
velocity of the BAL trough at small inclinations, and at a large
velocity at large inclinations. Thus, it may be used as an
indicator of the inclination angle of the system. Indeed, both
cases have been observed (O99).

The misalignment of the symmetric axes between the electron-scatterer
and resonance scatterer can also lead to PA rotation. However,
the physical driver for such misalignment is not clear. Furthermore,
if the size of the electron-scatterer is small and the optical depth of
the line is large, one would expect a constant PA rotation across
the BAL trough, which seems not a general case for BAL QSOs,
for non-rotating flows because the resonant scattered light dominates the
polarization in the trough. A large electron scatterer can produce
velocity-dependent PA rotation in the trough due to back-scatter
effect as discussed in the \S 4.1 because the scattered photons
from the reflection-symmetric sites will encounter the flow of
different velocities on the way to the observer. The large
electron scatterer will also produce a PA rotation without a jump
at the start velocity of the trough, just same as the axisymmetric
rotating outflow model with a LESR.

\subsection{Polarized Flux and Sub-troughs}
Several models predict distinct features, which reflect a special
velocity field or geometry of the outflow. Some of
these features are indeed observed in a number of BAL QSOs. Red
sub-trough in the polarized flux is predicted by models of large
$\theta_0$, and appears in the polarized spectra of several BAL
QSOs such as 0105-265, 0226-1024, 1333+2840, 1413+1143 (O99). The
observed sub-trough is usually narrower and shallower than the
primary one (blue), in agreement with our simulations. There is
also an emission like feature around $v=0$ (but no correspondent
CIII] emission feature) in the polarized flux in some of these
sources, such as 1413+1143. It should be noted that our model
predicts large PA rotation for these objects in general (Table 2).
Three out of four objects indeed show large PA rotations
(-10$^{\rm o}$ for 0226-1024, 10$^{\rm o}$ for 1333+2840 and
20$^{\rm o}$ for 1413+1143). No apparent PA rotation in 0105-265
may be an indication of low angular-momentum outflow or a large
optical depth to the resonant scattering in this object. The
appearance of the sub-trough in the polarized flux is ascribed to
the rotation of the STOKES $Q$ of the resonance scattering light
(Appendix A) across the absorption line profile. Direct evidence for
this comes from the polarization observation of 0043+0048: the
continuum is not polarized but the BAL trough is; the Stokes
parameters $Q$ is about 5\% of the total intensity around the
absorption trough, and changes its sign around zero velocity,
similar to our numerical results with a large $\theta_0$.

Ogle (1997) proposed that the red sub-trough in 0226-1024 can be
produced by polar an electron-scattering region with an equatorial
BALR. To investigate the feasibility of this scenario, we carry
out the simulation of resonant scattering for the case in which
the incident continuum is polarized in the direction perpendicular
to the symmetric axis of the resonant scatterer (PA$_c=90$$^{\rm
o}$), as produced by a polar electron-scatterer. Models
A0-12$^{\rm o}$, A0-45$^{\rm o}$, B0-33$^{\rm o}$-20$^{\rm o}$ and
B0-45$^{\rm o}$-30$^{\rm o}$ are used in this simulation. The
simulated polarized spectra are shown in Fig. \ref{com-pa90}. The
polarization of the primary trough is indistinguishable from those
with equatorial electron-scattering presented in \S 4.3 for all
models considered. This is expected because the polarized flux in
the trough is dominated by the resonantly scattered light for the
range of optical depth concerned. However, the profile to the red
side is very different for polar and equatorial scattering. These
results are consistent with the analysis presented in Appendix A.
For model A-12$^{\rm o}$ with PA$_c=90^{\rm o}$, the polarized
flux to the red side of the primary trough is also cancelled and
the sub-trough is broader than and blending with the primary
trough, which is not consistent with the observation. For model B
there is a peak to the red side of the trough, which is not
consistent with the observation too. Only in one case A0-45$^{\rm
o}$ viewed in $i=90^{\rm o}-\theta_0$, there is a separated
sub-trough to the red side of the primary trough, but the
sub-trough is blue-shifted and a peak appears to the red side of
it. So the model with large $\theta_0$ and PA$_c=0$$^{\rm o}$
should be a better explanation for the sub-trough in 0226-1024 and
other sources than that with PA$_c=90^{\rm o}$(or polar electron
scatterer).

Interestingly, 0932+5006 displays a polarized flux peak to red
side of the primary trough which has no corresponding emission
lines in the total flux. According to our analysis it can be
produced by a polar electron scattering region plus an outflow
with a geometry similar to the model B. But this is a rare case
among the 36 sources in O99. It must be mentioned that if the
electron scatterer covers the same sky as the BALR
for model B-45$^{\rm o}$-30$^{\rm o}$, according to the $Eq.$ 7
and 8 of Paper I, the scattered light by electrons is polarized in
the direction perpendicular to the axis of the scatterer. The
polarized flux distribution to the red side of the trough is same
as the result of model B-45$^{\rm o}$-30$^{\rm o}$ plus a polar
electron scatterer. So the source 0932+5006 can be interpreted by
this model, as well.

Another characteristics in polarized flux is noted by Ogle that in
some objects the troughs in polarized flux is more blueshifted
than that in the total flux. In our models with LESR the starting
velocity of the absorption in polarized flux is about
$-v_{\varphi,0}$ (if $v_{\varphi,0}>v_0$) or $-v_0$ (if
$v_{\varphi,0}<v_0$), if $v_{\varphi,0}>v_0$ it reproduces this
characteristic. Model B with LESR might produce $\sim$90$^{\rm o}$
PA swing at velocity $v<-v_{\varphi,0}$ in our simulation, it is
also found in several cases, for example 1212+1445 (CIV trough) and
1232+1325 (NV+Ly$\alpha$ trough). The two characteristics both
indicate that the electron scattering region and the resonant
scattering region are very close, or even coexist. Objects 1212+1445
and 1232+1325 are both low ionized BAL QSOs so the column of
outflow is large and consistent with our results.

\section{The Contribution of the Resonant scattered photons to the Broad
Emission Line}

Around half of the resonantly scattered photons in the BALR are absorbed
by the accretion disk, and the rest emerge as emission line photons. If
the BALR exists in all quasars, the contribution of the resonantly
scattered photons to be observed emission lines might be non-negligible
for non-BAL QSOs. In Fig. \ref{pf-nonBAL} we present the polarized flux
resonantly scattered into the line of sight of non-BAL QSOs, which appears
unique asymmetric, and thus may be used to test the presence of BALR in
non-BAL QSOs.

Their contribution is especially important for NV emission line,
which was used in the determination of the metal abundance in
quasars (Hamann \& Ferland 1992; 1993). HKM93 considered this
problem in detail. Our treatment is different from theirs in
serval aspects. First of all, we perform Monte-Carlo simulations
to calculate the radiative transfer instead of using an escape
probability approximation. Second, we consider a much larger range
of the scattering optical depth in comparison with theirs, as
required by observations. Third, we consider the rotation velocity
of the flow and use a different radial velocity-law. Finally, we
consider the absorption of the scattered photons by the accretion
disk. In this section we mainly consider model A-12$^{\rm o}$.

\begin{figure}
\epsscale{0.45}\plotone{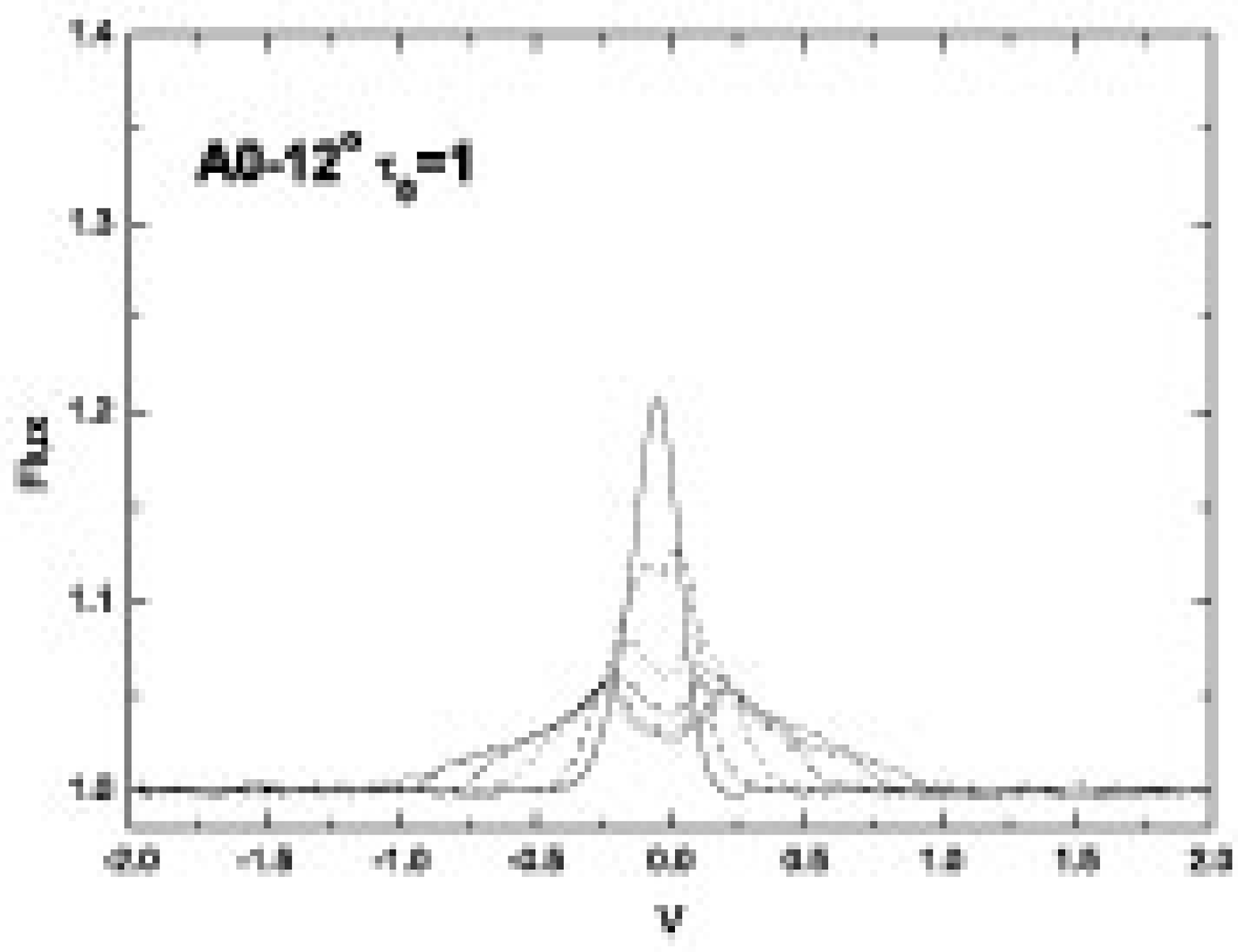}\plotone{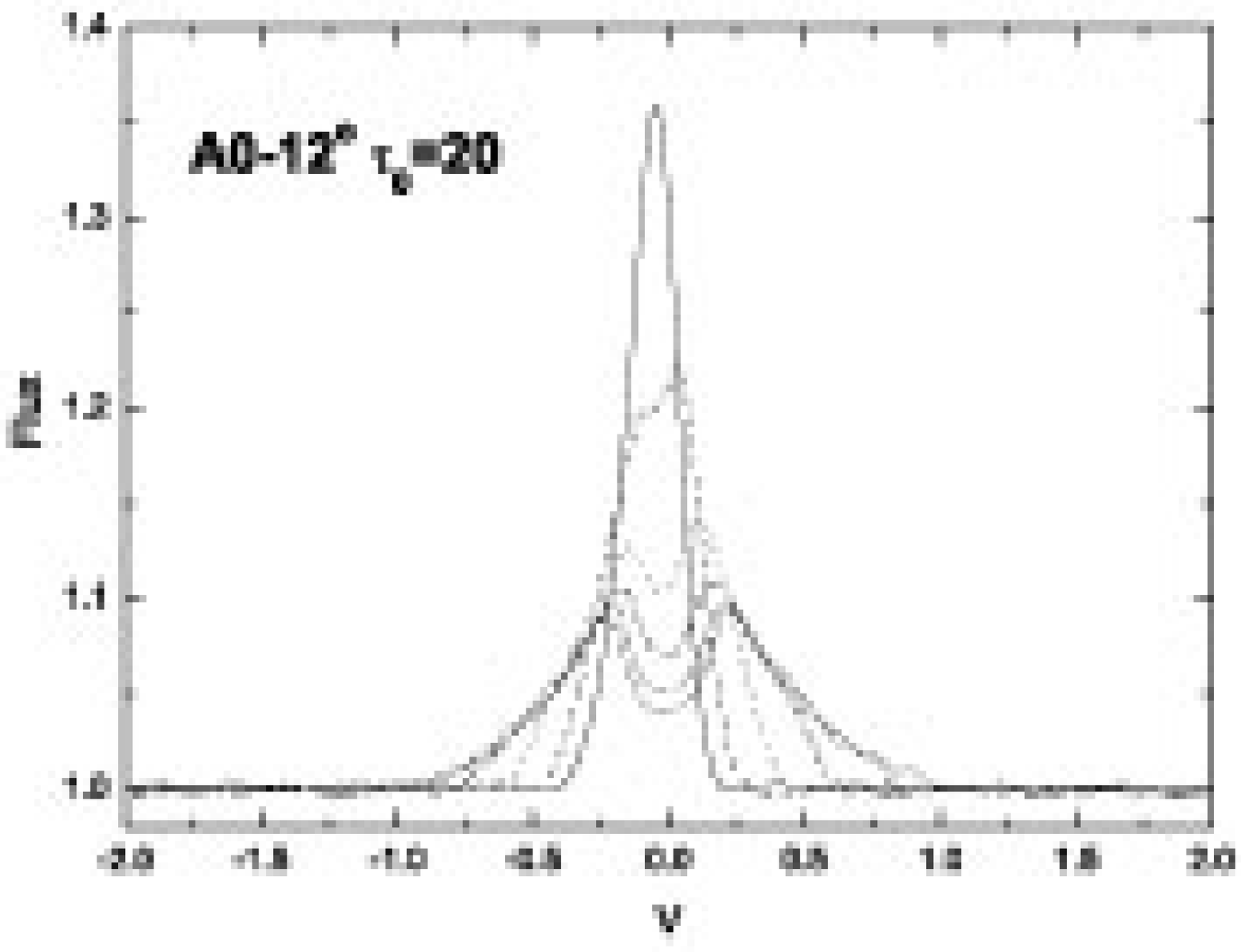}
\plotone{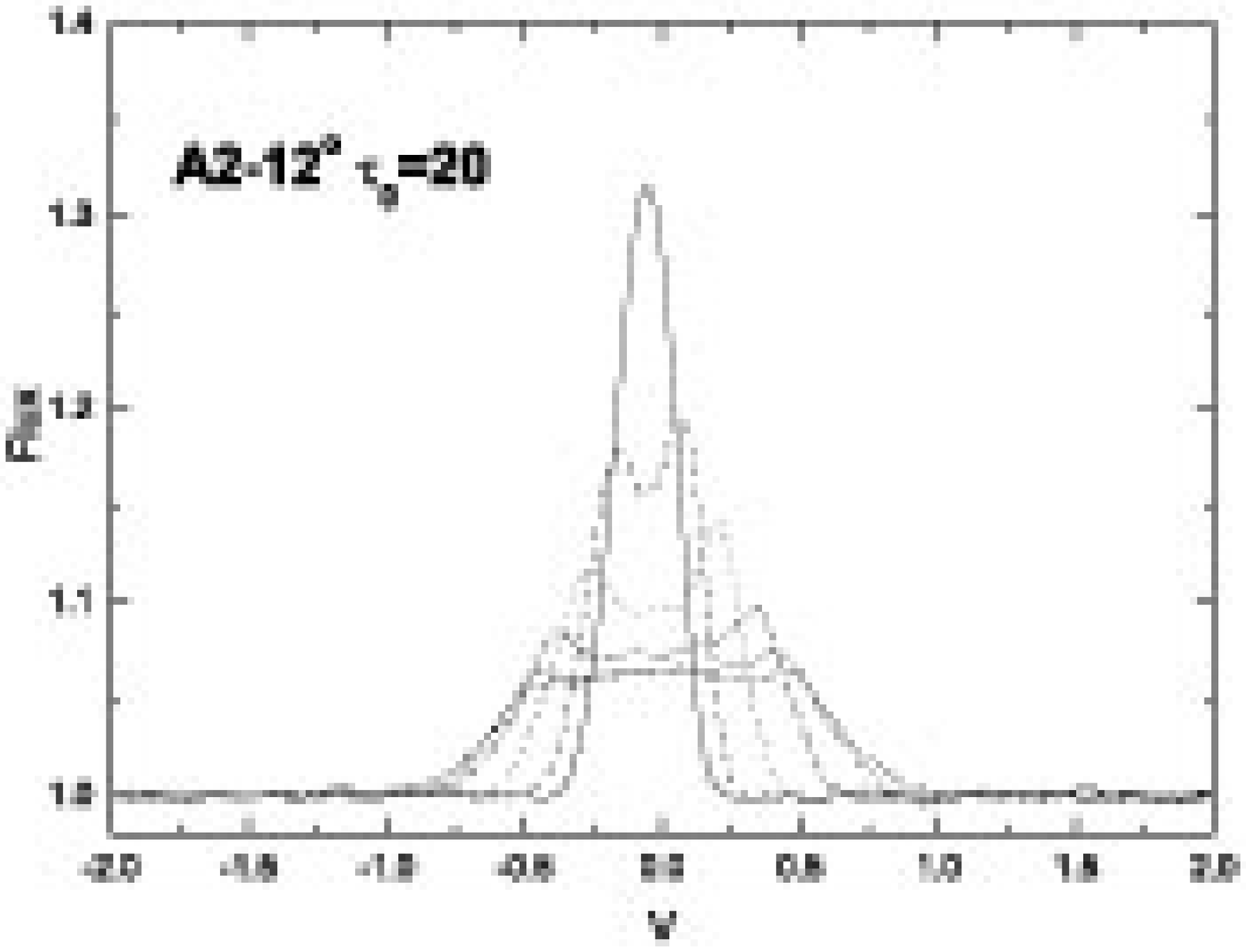}\plotone{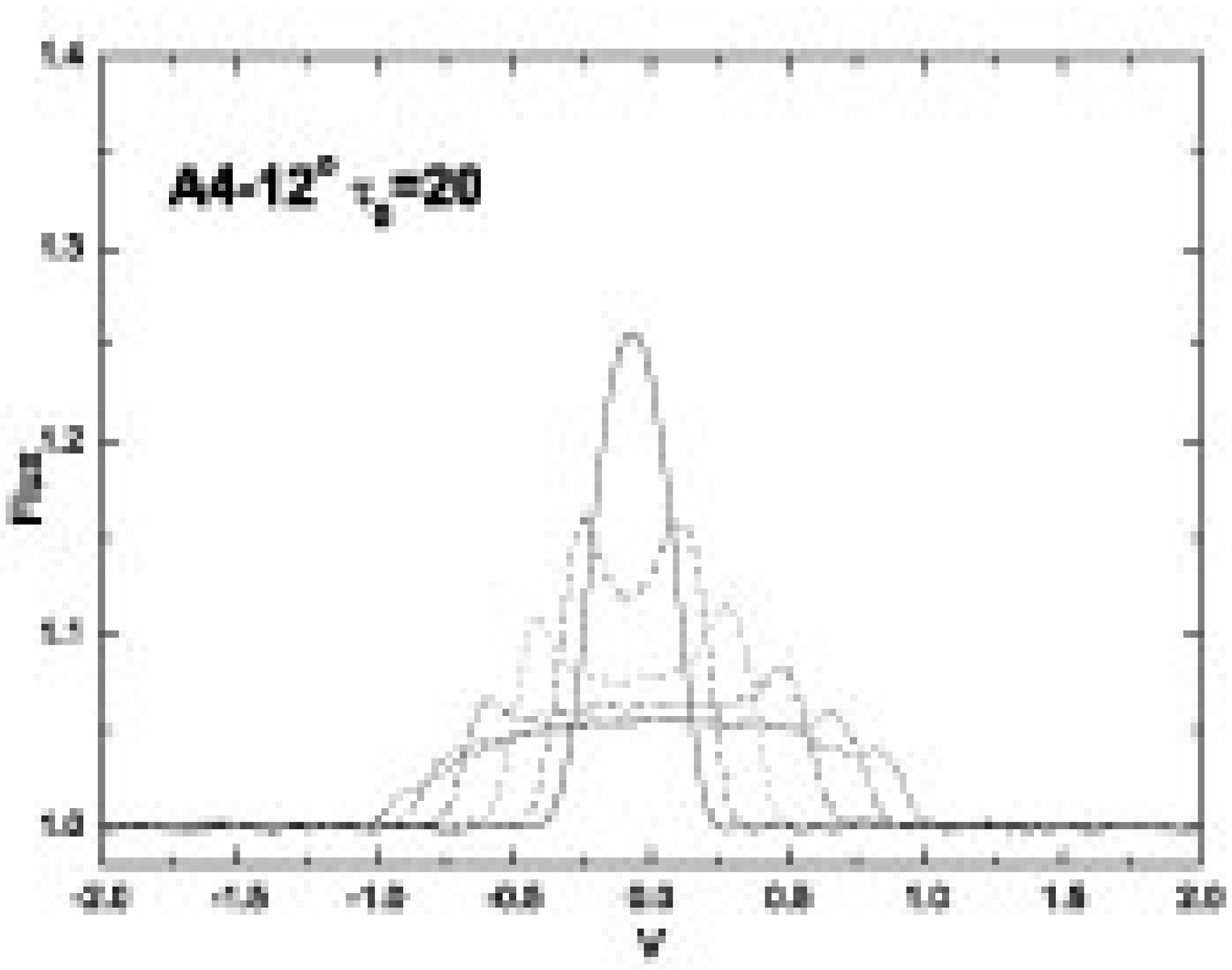} \caption{Scattered line
profile (continuum has been normalized to unity)
   viewing at 6 different inclinations: $i\sim$ 0\deg, 16\deg, 27\deg,
   42\deg, 58\deg and 77\deg for non-BAL QSOs. Parameters of the flows are:
   $\tau=1$ and $q=0.0$ in top-left panel, and $\tau=20$ and $q=0.0$
   in the top-right panel, and $\tau=20$ and $q=0.5$ in bottom-right
   panel, $\tau=20$ and $q=1.0$ in the bottom-right panel. }\label{sca-em}
\end{figure}

For a pure radial outflow, the line profile of the scattering
emission is qualitatively similar to that of HKM93 (Fig.
\ref{sca-em}, the upper two panels and figure 7, 8 of HKM93).
Note due to the disk absorption, the contribution of the scattered
photons only come from the facing half of the BALR, and the
line profile is slightly blue-shifted relative to the HKM93 (Fig.
\ref{sca-em}).
When the rotation is significant, the profiles are much
different (Fig. \ref{sca-em}, lower two panels): at
intermediate inclinations the two peaks in the profile are further
separated because the scattered photons preferably escape in the
rotational direction, along which the velocity gradient is large; at high
inclinations, the peaks disappear and the line profiles become flatter
and the half width of the "platform" is about $q\times v_\infty=
v_{\varphi,0}$.

\begin{figure}
\epsscale{1}\plotone{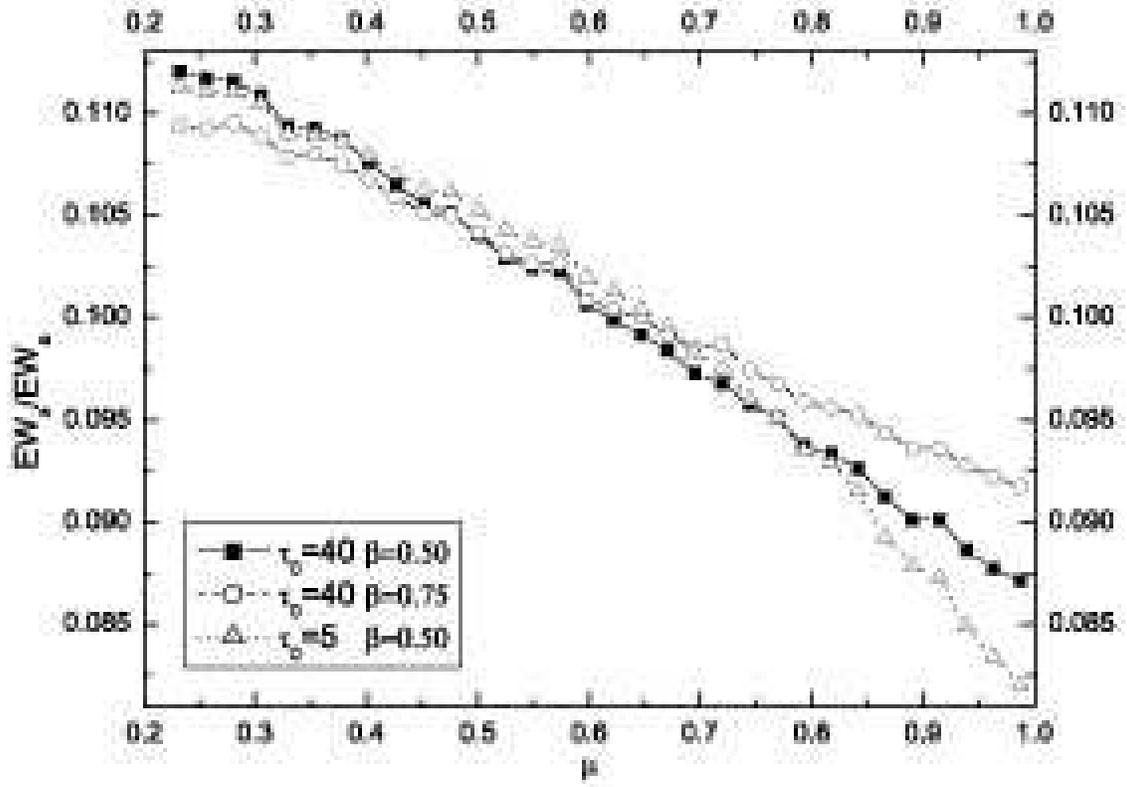} \caption{The equivalent width ratio
of the scattering emission to the BAL
  for model A-12$^{\rm o}$ as a function of $\mu=\cos i$.}\label{ew-u}
\end{figure}

\begin{figure}
\epsscale{0.45}\plotone{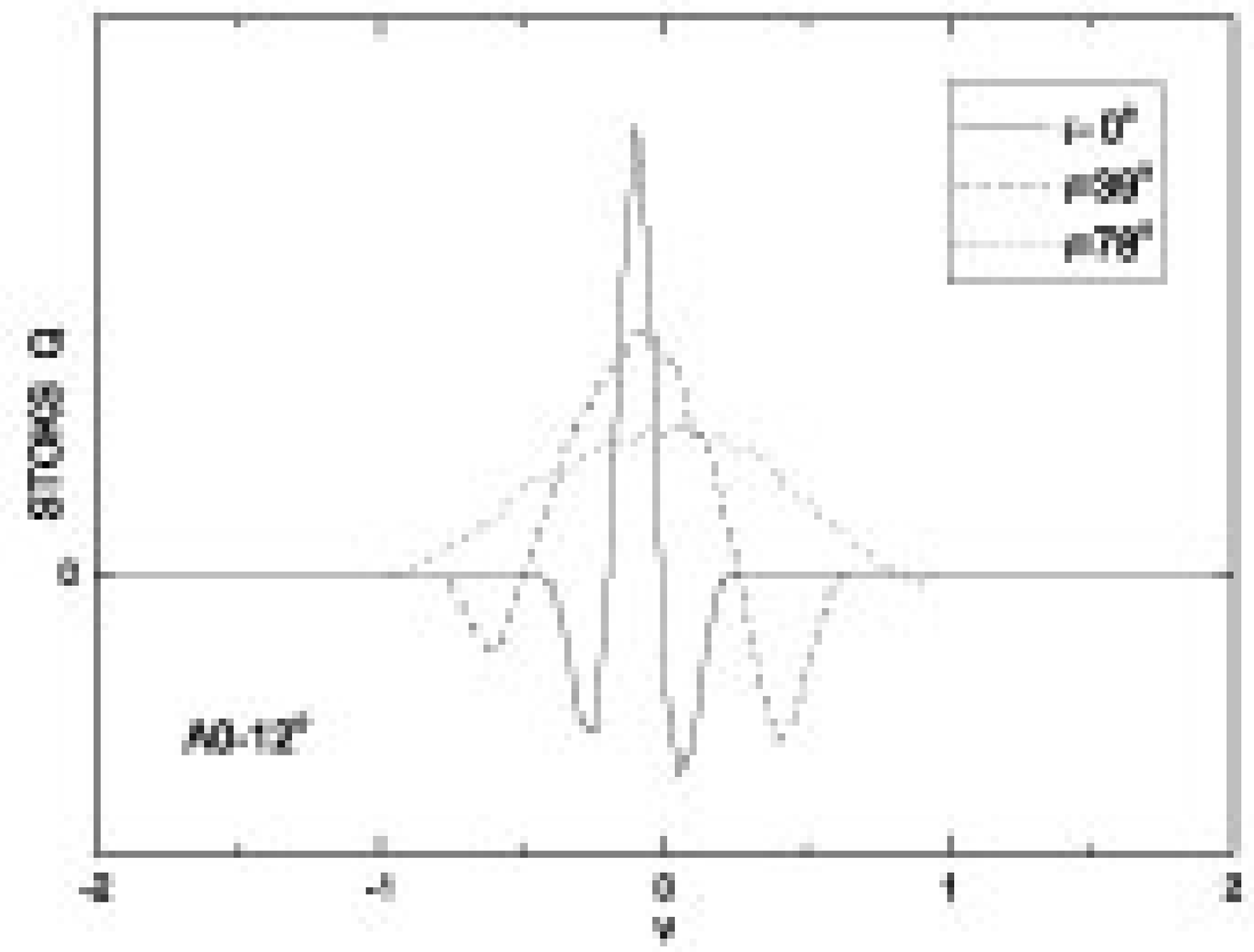}\plotone{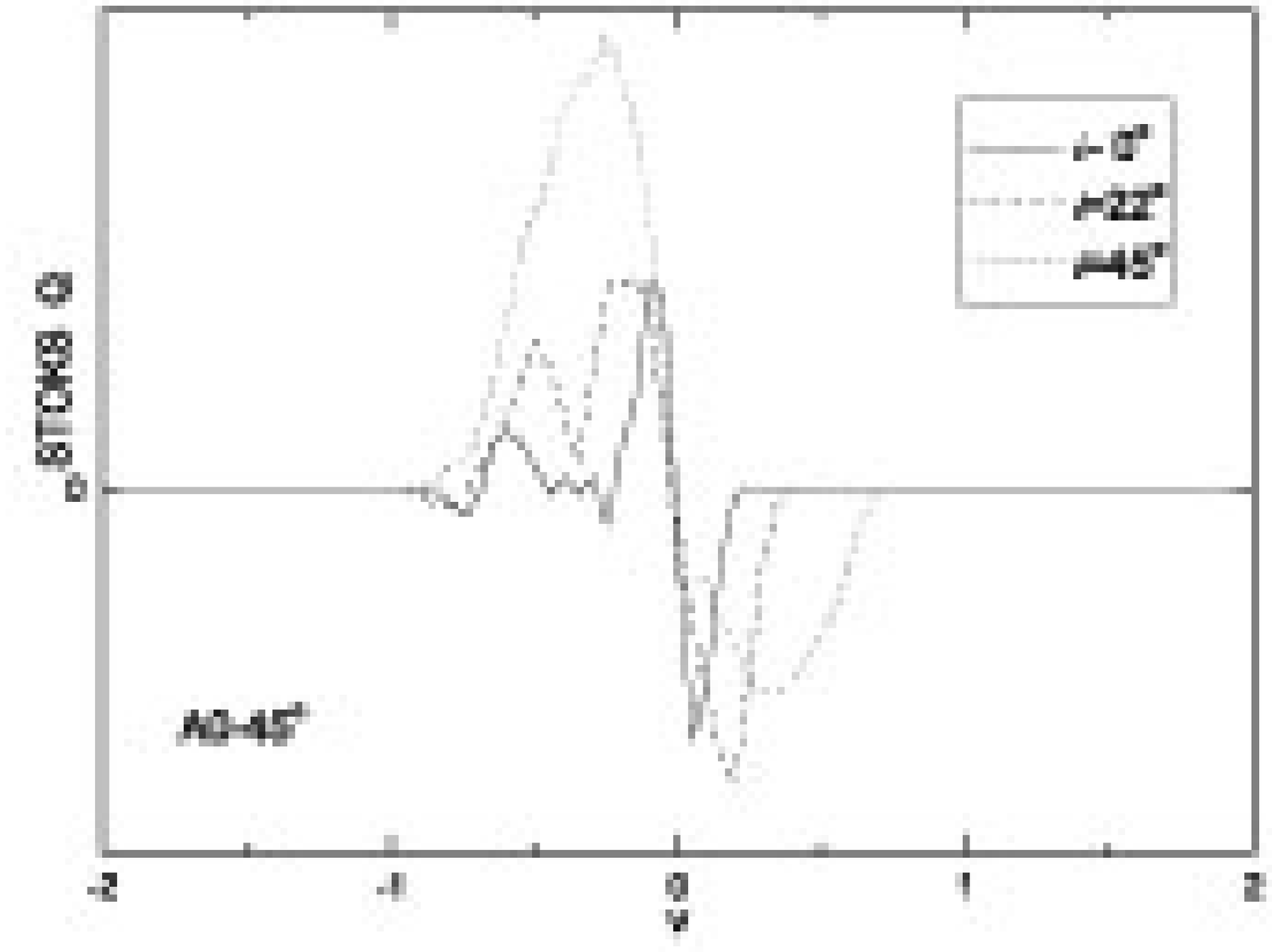}
\plotone{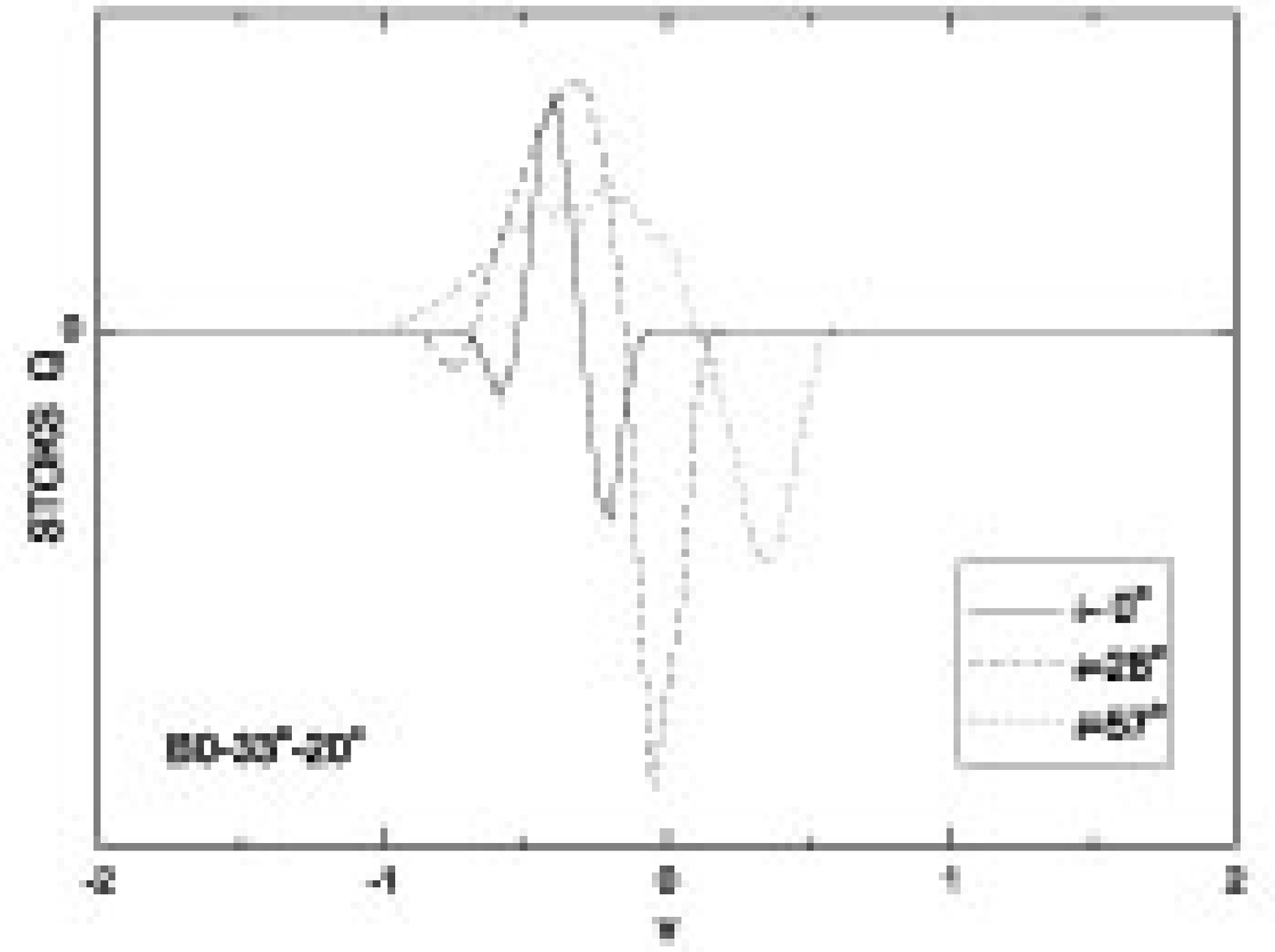}\plotone{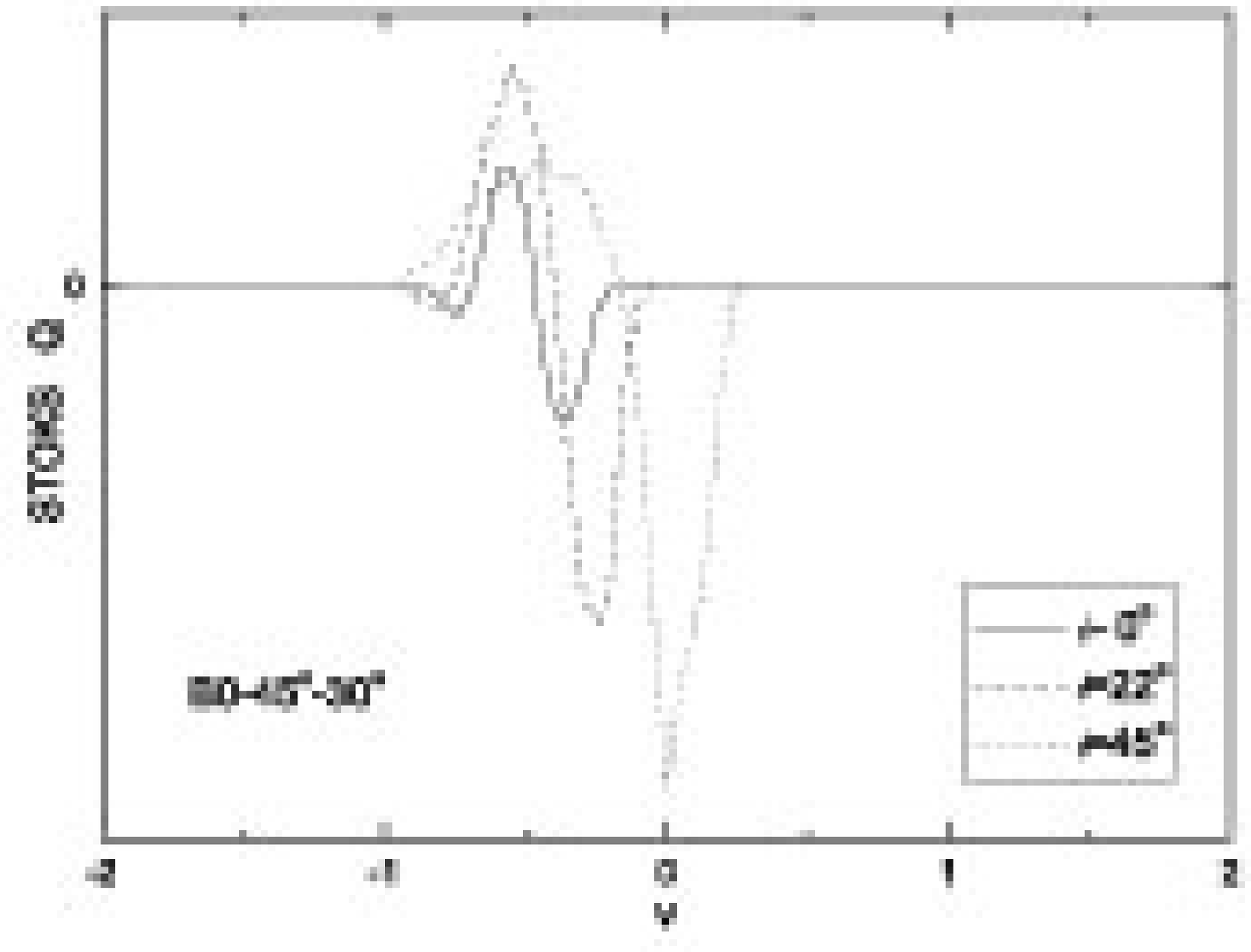} \caption{The simulated
polarized flux, resonantly scattered by the BALR to different
inclinations. Models of the BAL outflow are labelled in the figure
and $\tau_0=5$. }\label{pf-nonBAL}
\end{figure}

The scattering emission is stronger in the equatorial direction,
at least for $\beta=0.5$ and $\beta=0.75$ (Fig. \ref{ew-u}). For
model A4-12$^{\rm o}$ with the $\tau_0=40$, we get a maximum EW (for
non-BAL QSOs) of the scattering emission $w_{sl}$ which is about
11\% of the corresponding BAL EW. The fraction increases to
18.6$\%$ for model A4-24$^{\rm o}$. The scattering emission of NV
1240 is even stronger because of the scattering of the strong
Ly$\alpha$ emission line. For example, in 0105-265 nearly half of
Ly$\alpha$ emission line, of which the EW is $\approx250$\AA, is
scattered by the NV ions (O99). According to the ratio noted
above, the total EW of the scattering NV emission would be
$(EW_{{\mathrm Ly}\alpha}/2+EW_{BAL})\times11\%=20.8$\AA~for A4-12$^{\rm
o}$, and 35.2\AA~for A4-24$^{\mathrm o}$ if the BAL extends to 2
$\times$ 10$^4$ km~s$^{-1}$. This is significant considering the
observed EWs of NV emission line is only in the range of 9.2\AA~
to 44.8\AA~ (Ferland et al. 1996). Therefore, a strong NV may not
indicate a high metal abundance but a large covering factor of the
BALR. If this is indeed the case, we also expect that the EW of
NV$\lambda$1240 be correlated with continuum polarization.

HKM93 used $W_\lambda(em)/W_\lambda(abs)$ of the BAL QSOs to set
an upper limit of the covering factor, about 0.2. Here we point
out that if the disk absorption is considered, the upper limit
should bounce up to 0.4. Following HKM93's calculation (their
$Eq.$ 1), we estimate the effective "covering factor" from the
simulated total flux spectrum for each model. As expected, we find
that the effective "covering factor" (using $Eq.$ 1 of
HKM93) is about half the solid angle subtending by BALR. For the
models in the previous section, we obtained a "covering factor" of
0.34$\sim$0.40 for model A0-45$^{\rm o}$, 0.26$\sim$0.28 for model
A0-30$^{\rm o}$, 0.115 for model A0-12$^{\rm o}$, 0.06$\sim$0.08
for model B0-45$^{\rm o}$-30$^{\rm o}$ and 0.10 for model
B0-33$^{\rm o}$-20$^{\rm o}$.

Note in passing, polarization observations of resonant lines in
non-BAL quasars should provide a good test to the unification
model of BAL and non-BAL. The polarization of the scattering lines
show unique profile across the line profile, and depends strongly
on the inclinations and model parameters (Fig 26). In particular,
the subtending angle of the flows is also directly relative to
amount of scattering light, thus the polarization degree. The
observations would be preferably to be done for NV-strong quasars
if the interpretation of strong NV emission as due to resonance
scattering is correct. Thus the observation, in turn, provides a
check for the scenario.

\section{Summary}

Polarization provides rich information on the structure and
kinematics of BALR in QSOs, which is complementary to those
derived from absorption lines, which only depend on the physical
condition and kinematics of gas on the LOS. To extract this
information, we carry out extensive Monte-Carlo simulations of
electron and resonance scattering process in the BALRs. Both
singlet and doublet transitions are considered for radial outflows
of two different geometries: equatorial outflows and hollow-conical
outflows with and without rotational velocities.

In an axisymmetric scatterer model, PA rotation in the absorption
trough can be produced only when the outflow carries angular
momentum. The PA rotation increases with the angular velocity as
well as the subtending angle of the flow. In order to produce a PA
rotation $>$ 10\deg observed in a few BAL QSOs, subtending angle
of the outflow should be larger than 25\deg. Similar requirement
is imposed to explain the red sub-trough in the polarized flux observed
in some BAL QSOs.

Resonant scattering of doublet transition produces much lower
polarization at large optical depths (about 6\% for $\tau_0=5$,
$\beta=0.5$) for an accelerating outflow.
Large
($>$10\%) polarization in the absorption trough detected in a few
BAL QSOs may indicate that the optical depth to the
resonant scattering is at most moderate, otherwise resonant
scattering of $T_{1/2}$ transition will erase all polarization.
A slowly-accelerating
model can produce larger polarization. If the observed
high-velocity outflow in X-ray is real, the flow is likely
decelerating, which will produce much larger polarization.

A large electron scattering region can also produce larger
polarization. The jump which appears at the starting velocity of the
absorption trough in PA rotation and polarized flux in models
with a SESR does not appear in the model with a LESR. This
characteristic may be an important indicator to distinguish
different electron scattering models. We find that LESR models can
explain that the absorption troughs are more blueshifted in polarized
flux, relative to that in the total flux, and PA swings in the troughs,
$\sim 90^{\rm o}$ relative to continuum, in some objects.

We show that the resonantly scattered light will contribute a
significant part of NV in some QSOs and can give rise to anomalous
strong NV lines in these QSOs. A correlation between the EW of
NV$\lambda$1240 with the continuum polarization is expected. We
propose that the polarized flux and the PA rotation of the
scattering emission can be used to test the presence of BALR in
non-BAL QSOs. Further observations with large telescopes should
allow us to extract the important information about the flow
geometry and kinematics.

\acknowledgements We thank the referee for useful suggestions.
This work was supported by Chinese NSF through
NSF 10233030 and NSF 10573015, the Bairen Project of CAS.

\newpage
\appendix{Appendix A: Notes on the red-trough in the polarized flux}

We denote the Stokes parameters $Q$ and $U$ of the continuum as
$Q_c$ and $U_c$. For simplicity, we choose $U_c=0$ and
$|Q_c|=I^c_{pl}$, where $I^c_{pl}$ is the polarized flux of the
continuum. If $Q_c>0$ the PA of the continuum polarization (hereafter
$\mathrm{PA}_c$) is 0$^{\rm o}$ otherwise $\mathrm{PA}_c=90$$^{\rm
o}$. We assume that PA of the line scattered photon is
PA$_l$ so the $U_l=Q_l\tan($2PA$_l)$. The Stokes
parameters of the total flux are $Q_T=Q_c+Q_l$, $U_T=U_c+U_l$; the
total polarized flux $I^T_{pl}$ reads,

\begin{equation}
I^T_{pl}=\sqrt{U_T^2+Q_T^2}=\sqrt{Q_l^2(1+\tan^2(2\rm{PA}_l))+Q_c^2+2Q_cQ_l}\label{app1}
\end{equation}
\\
From this equation we find that $I^T_{pl}<I^c_{pl}$ if

\[ \left\{
           \begin{array}{ll}
               0>Q_l>-\frac{2Q_c}{1+\tan^2(2\rm{PA}_l)} &  (\mathrm{PA}_c=0^{\rm o}) \\
               0<Q_l<-\frac{2Q_c}{1+\tan^2(2\rm{PA}_l)} & (\mathrm{PA}_c=90^{\rm o})
            \end{array}
            \right.
 \]
\\
When $Q_l\times Q_c<0$ and $|Q_l|\lesssim |Q_c|$, one obtains
$I^T_{pl}<I^c_{pl}$, i.e, an absorption trough appears in the
polarized flux. Since the total flux equals to the $I_c+I_l>I_c$
the feature is not shown in the total flux. On the other hand, if
$Q_l \times Q_c>0$, one yields $I^T_{pl}>I^c_{pl}$. In the rest of
this appendix we will obtain the relationship between $Q_l$ and
$Q_c$ for different models.

For simplicity, we consider single-scattering of the unpolarized
incident spectrum by singlet transition. The density matrix of the
incident continuum is

\begin{equation}
\rho^i_{11}=\rho^i_{22}=1/2;\rho^i_{12}=\rho^i_{21}=0
\end{equation}

Now consider an outflow between [90$^{\rm o}-(\theta_0-d\theta_0)$,
90$^{\rm o}$$-\theta_0$]. For light from an incident direction
$(\theta_i,\varphi_i)$ that is scattered into the direction
$(\theta_o,\varphi_o)$, the density matrix can be written as (see $Eq.$ 1
\& 2 of Paper I):

\begin{equation}
\rho^o_{11}\propto\frac{1}{2}[\sin^2\theta_0\cos(\varphi_i-\varphi_o)+\cos^2\theta_0]^2+\frac{1}{2}\sin^2\theta_0\sin^2(\varphi_i-\varphi_o)
\end{equation}

\begin{equation}
\rho^o_{22}\propto\frac{1}{2}\sin^2\theta_0\sin^2(\varphi_i-\varphi_o)+\frac{1}{2}\cos^2(\varphi_i-\varphi_o)
\end{equation}
\\
The Stokes parameter $Q$ of the scattered light reads

\begin{equation}
Q_l(\varphi_i,\varphi_o)=\rho^o_{11}-\rho^o_{22}\propto\frac{1}{2}[\sin^2\theta_0\cos(\varphi_i-\varphi_o)+\cos^2\theta_0]^2-\frac{1}{2}\cos^2(\varphi_i-\varphi_o)
\end{equation}

It is easy to prove
\begin{equation}
 Q_l \geq 0 \;\; {\mathrm for}\;\; -90^{\rm o}<\varphi_i-\varphi_o<90^{\rm
o}
\end{equation}
 and $Q_l<0$, when
\begin{equation}
180^{\rm
o}+\csc\frac{1-\sin^2\theta_0}{1+\sin^2\theta_0}>\varphi_i-\varphi_o>180^{\rm
o}-\csc\frac{1-\sin^2\theta_0}{1+\sin^2\theta_0}
\end{equation}

In an outflow model, red-shifted photons are escaped from the
portion with 90$^{\rm o}<\varphi_i-\varphi_o<270^{\rm o}$ and the
blue-shifted ones from the -90$^{\rm
o}<\varphi_i-\varphi_o<90^{\rm o}$. According to $Eq.$ A6, the
blue-shifted photons always have $Q_l\geq0$. However, the
redshifted scattered light may have negative $Q_l$ following $Eq.$
A7. The portion of the flow that produces negative $Q_l$ increases
with $\theta_0$. It is small for small $\theta_0$ and reachs half
of backward flow (135$^{\mathrm o} <\varphi_i-\varphi_o<
225^{\mathrm o}$) for $\theta_0\sim25^{\mathrm o}$. Consequently,
for a small $\theta_0$ the total $Q_l$ of red-shifted scattered
photons is larger than 0. If $Q_c$ is also positive, an excess
polarized flux across the emission line will be seen; otherwise,
an absorption trough is observed (Fig. 22).  For a large
$\theta_0$, redshifted scattered light has negative $Q_l$, a
sub-trough to the red side of primary trough will be observed when
$Q_c$ is positive.

\newpage


\end{document}